\documentclass[9pt,english]{extarticle}
\usepackage{amsmath} 
\usepackage{epstopdf}
\usepackage{graphicx} 
\usepackage{balance} 

\usepackage{epstopdf}

\usepackage{multirow} 
\usepackage{array}

\usepackage{wrapfig} 
\usepackage[utf8]{inputenc} 
\usepackage{clrscode} 
\usepackage{listings} 
\usepackage{booktabs}

\usepackage{subfigure} 
\usepackage{hyperref}
\usepackage{gensymb}
\usepackage{tikz}
\usepackage{verbatim}
\usepackage{scalefnt}
\usetikzlibrary{arrows}
\usetikzlibrary{positioning}
\usetikzlibrary{decorations.markings}
\usetikzlibrary{decorations.pathreplacing}

\usepackage[utf8]{inputenc}
\usepackage{amsmath}
\usepackage{amsfonts}
\usepackage{amssymb}
\usepackage{amsthm}
\usepackage{graphicx}
\usepackage{color}
\usepackage[ruled,linesnumbered]{algorithm2e}
\usepackage{caption}
\usepackage{url}
\usepackage{booktabs}

\usepackage{palatino}

\newcommand{\ra}[1]{\renewcommand{\arraystretch}{#1}}

\DeclareMathAlphabet{\mathpzc}{OT1}{pzc}{m}{it}

\newcommand{\ignore}[1]{}

\usepackage[left=2cm,right=2cm,top=2cm,bottom=2cm]{geometry}

\usepackage{palatino}

\begin{document}

\title{\Huge{Location Estimation Using Crowdsourced Geospatial  Narratives}}

\author{
  Georgios Skoumas\\
  School of Electrical and Computer Engineering, NTUA \\
  \texttt{gskoumas@dblab.ece.ntua.gr}
  \and
  Dieter Pfoser\\
  Dept. of Geography and Geoinformation Science, GMU \\
  \texttt{dpfoser@gmu.edu}
  \and
  Anastasios Kyrillidis\\
  School of Computer and Communication Sciences, EPFL \\
  \texttt{anastasios.kyrillidis@epfl.ch}
}

\maketitle

\begin{abstract}

The ``crowd'' has become a very important geospatial data provider. Subsumed under the term Volunteered Geographic Information (VGI), 
non-expert users have been providing a wealth of quantitative geospatial data online. 
With spatial reasoning being a basic form of human cognition, narratives expressing geospatial experiences, e.g., travel blogs, would provide 
an even bigger source of geospatial data. Textual narratives typically contain qualitative data in the form of objects and spatial relationships.
The scope of this work is $(i)$ to extract these relationships from user-generated texts, $(ii)$ to quantify them and $(iii)$ to reason about object 
locations based only on this qualitative data. We use information extraction methods to identify toponyms and spatial relationships and to formulate a 
quantitative approach based on distance and orientation features to represent the latter. Positional probability distributions for spatial relationships 
are determined by means of a greedy Expectation Maximization-based (EM) algorithm. These estimates are then used to ``triangulate'' the positions of unknown object locations. 
Experiments using a text corpus harvested from travel blog sites establish the considerable location estimation accuracy of the proposed approach.
\end{abstract}

\section{Introduction} 
\label{sec:introduction}

User-contributed content has benefited many scientific disciplines by providing a wealth of new data sources. 
In the geospatial domain, authoring content typically involves quantitative, coordinate-based data. 
While technology has helped a lot to facilitate geospatial data collection, e.g., all smart phones are equipped with 
GPS positioning sensors, yet authoring quantitative data requires specialized applications (often part of social media platforms) and/or specialized knowledge, e.g., Openstreetmap\footnote{\url{https://www.openstreetmap.org/}}.
This fact hinders the widespread adoption of VGI as an even bigger, large-scale geospatial data source. 

The broad mass of users contributing content on the Internet are much more comfortable using \emph{qualitative information}. 
People typically do not use coordinates to describe their spatial experiences (trips, etc.), but rely on qualitative concepts in the form of toponyms (landmarks) and spatial relationships (near, next, etc.). 
Crowdsourcing qualitative geospatial data is thus more challenging since the mental model of space is actually very different from the representation that is used to record datasets.

Utilizing qualitative geospatial data is typically based on trying to quantify it. One of the challenges here is the uncertainty associated with the data. The same concept (near) might be interpreted differently by the various users. 
As an example, consider the following narrative.
\textit{``The best pita place in Greece is \textbf{next to} the Monastiraki Metro Station in Athens.''}
In this case, we want to quantify what people imply when they say \textit{``next to''}. Being able to do so, might allow us to actually discover the ``best pita place in Greece''. 
%
Eventually, by collecting more observations that mention the ``best pita place in Greece'' using qualitative spatial information, i.e., spatial relationships tying the place to known locations, 
we will be able to refine the unknown location and, thus, locate places that otherwise could not be geocoded. 
\vspace{12pt}

To this end, we consider the following problem:

\vspace{12pt}

\noindent \textsc{Problem:}
\textit{Given a set of objects $P_{K}$ with a-priori known coordinates in space, a set of objects $P_{U}$ whose exact
positions are unknown, and a set of observed spatial relationships $R$ between objects of set $P_{U}$ and objects of set $P_{K}$, find probabilistic estimates 
of the positions of objects in the set $P_{U}$, based on their observed spatial relationships $R$ with objects in the set $P_{K}$.}
\vskip.05in
\vspace{5pt}

Even though the above formalization is very comprehensive, the problem contains high uncertainty, especially when the source of spatial information is user-contributed. 
To achieve high accuracy location estimates, our approach follows a probabilistic path, where we quantify qualitative relations as probability measures. Essentially, using textual narratives, 
we observe point-of-interest (POI) pairs that are linked together by a spatial relationship. 
Assuming that both locations are known, each observation is roughly quantified using a spatial feature vector comprising distance and orientation. 
We use information extraction methods to identify those toponyms and spatial relationships in texts.
A greedy Expectation Maximization-based (EM) method is used to train a probability distribution, which represents the quantified spatial relationships under a probabilistic framework. 
Given a specific spatial relation, it provides a set of random variables (spatial feature vector) that have certain probability density functions (PDFs) associated with them, for a specific 
spatial relation. These positional PDFs are then used to ``triangulate'' the positions of unknown POI locations. The more observations we have with respect to an unknown location, 
the preciser we will be able to reason 
about the POI's unknown location. 
Actual location estimation experiments using textual narratives from travel blogs establish the validity and quality of the proposed approach.



The outline of the remainder of this work is as follows. Section~\ref{sec:related_work} discusses related work. Section~\ref{sec:contribution} discusses the specific qualitative data involved and introduces the 
spatial feature vectors used for quantification, while Section~\ref{sec:modeling} introduces the tools necessary to derive quantification in the form of PDFs for the spatial 
relationships. Section~\ref{sec:evaluation} validates the proposed approach using synthetic and real world location estimation scenarios.
Finally, Section~\ref{sec:conclusions} presents conclusions and directions for future work.

\section{Related Work} 

\label{sec:related_work} 
Work relevant to this research includes: $(i)$ extraction of semantic or especially spatial relations from natural language expressions, $(ii)$
qualitative modeling of spatial relations and its application to spatial databases, and, $(iii)$ quantitative modeling of spatial knowledge.

The extraction of qualitative spatial data from texts requires the utilization of efficient natural language processing (NLP) tools to automatically extract and map phrases to spatial relations. 
In the past, extraction of \emph{semantic} relations between entities in texts is developed in \cite{bunescu2006subsequence, ReVerb, wanderlust09, sonex, Zelenko}, while extraction of \emph{spatial} 
relations between entities in texts is analyzed in \cite{spatialrolelabel, extrir, svmextr}. 
While the above works constitute a good match in our developments for spatial relationship extraction from texts, we intentionally designed our specialized qualitative spatial data mechanism 
that better fits into the particularity, e.g. noisy crowdsourced data, of the relation extraction part of our problem. 
Formal methods for qualitative representation of spatial relationships based on mathematical theories of order are presented in \cite{egenhof2, egenhof1, egenhof3, Kainz}.
Their applicability on spatial database systems and some key-role technical concepts are coherently discussed in \cite{Guting, Papadias94theretrieval}. 
Qualitative representation of spatial knowledge is discussed in \cite{chorochron2, Papadias} where the authors identify the common concepts of the qualitative representation and processing of spatial knowledge. 
In this work, we bridge the gap between qualitative and quantitative data by quantifying qualitative spatial relationships extracted from user generated texts.

Recent research on\emph{ quantitative representation of spatial knowledge} has been conducted in relation to situational awareness systems, robotics, and image processing. 
Modeling uncertain spatial information for situational awareness systems is discussed in \cite{Kalashnikov1, Kalashnikov2}. 
The authors  propose a Bayesian probabilistic approach to model and represent uncertain event locations described by human reporters in the form of free text. 
Estimation of uncertain spatial relationships in robotics is addressed in \cite{roboticsquantrep}. 
%
A probabilistic algorithm for the estimation of distributions over geographic locations is proposed in \cite{im2gps} where 
a data-driven scene matching approach is used in order to estimate geographic information based on images. 
Image similarity based on quantitative spatial relationship modeling is addressed in \cite{Wang}
while, in \cite{fuzzyquantrep}, a fuzzy decision tree algorithm is proposed to formalize spatial relations between linear objects.
Finally, in \cite{skoumas}, the authors introduce a basic Expectation Maximization/Gaussian Mixture Model approach to quantify qualitative spatial data manually extracted from texts. 
In this work, we provide a framework for the automatic extraction of qualitative spatial data from texts and we introduce a location estimation 
method based on spatial relation fusion. This bridges the gap between qualitative and quantitative representation of spatial relations using efficient machine learning techniques
and arrives at an actual text-to-map application: starting from extraction, moving to modeling and finally to location estimation. 




%
\section{Qualitative Spatial Data}
\label{sec:contribution}


This section highlights our approach on qualitative data extraction from texts and presents a model for representing spatial relationships based on distance and orientation measures.

\subsection{Textual Narratives as Data Source}
\label{subsec:data1} 
Crowdsourced narratives are likely to contain spatial information, if we focus on text that is related to spatial experiences.
In this work, we choose travel blogs as a rich potential geospatial data source. 
This selection is based on the fact that people tend to describe their experiences in relation to their trips and places they have visited, which results in ``spatial'' narratives. 
To gather such data, we use classical Web crawling techniques \cite{Drymonas} and compile a database\footnote{Available upon request} consisting 
of 120,000 texts, obtained from travel blogs\footnote{TravelBlog, TravelJournal, TravelPod}.

\subsection{Spatial Relations}
\label{subsec:data2} 

Obtaining qualitative spatial relations from text involves the detection of $(i)$ spatial objects, i.e., Points-of-Interest (POIs) or toponyms and $(ii)$ spatial relationships linking the POIs.
The employed approach involves geoparsing, i.e., the detection of candidate phrases, and geocoding, i.e., linking the phrase/toponym to actual coordinate information. 

Using the Natural Language Processing Toolkit (NLTK) (cf. \cite{nltk}), 
which is a leading platform to analyze raw natural language data,  
we managed to extract 500,000 POIs from the text corpus.  
For the geocoding of the POIs, we rely on the GeoNames\footnote{http://www.geonames.org/}  geographical gazetteer data, which covers all countries and contains over ten million place names and their coordinates. 
This procedure associates (whenever possible) geographic coordinates with POIs found in the travel blogs, using string matching based on the Levenshtein string distance metric (cf. \cite{levenst}). 
Using the GeoNames gazetteer we managed to geocode about 480,000 out of the 500,000 extracted POIs.

%

Having identified and geocoded the spatial objects, the next step is the extraction of qualitative spatial relationships. 
%
As mentioned is Section~\ref{sec:related_work}, the extraction of spatial relations between entities in text is a hard NLP problem. REVERB (i.e. \cite{ReVerb}) and EXEMPLAR (i.e. \cite{sonex}) are the state of the art
available software tools for the extraction of semantic relations between identified entities in texts. The main drawback of these approaches
is that they are not trained for spatial relations specifically. Moreover, we observe that they might perform poorly when applied with 
a noisy crowdsourced dataset. 


%
We address this NLP challenge by implementing a spatial relation extraction algorithm based on NLTK \cite{nltk} components in combination with predefined strings and syntactical patterns. 
More specifically, we define a set of language expressions that are typically used to express a spatial relation in combination with a set of syntactical rules. The use of both syntactical  
and string matching reduces the number of false positives considerably. As an example, consider the following phrase. \textit{``\textbf{Deutsche Bank} invested 10 million dollars \textbf{in} \textbf{Brazil}.''}. 
Here, a simple string matching solution would extract a triplet of the form (Deutche Bank, in, Brazil), which is a false positive. In our approach, the use of predefined syntactical
patterns avoids this kind of mistakes. On the other hand, for the phrase \textit{``\textbf{Deutsche Bank} invested 10 million dollars \textbf{in} \textbf{Rio de Janeiro}, which is \textit{``\textbf{in Brazil}}.''}  
our algorithm would extract a triplet of the form (Rio de Janeiro, in, Brazil) which is a true positive.

In Table~\ref{table:pre_recal}, we show a spatial relation extraction task on a small dataset of 300 
annotated crowdsourced spatial relations. Both our method and EXEMPLAR perform better than REVERB in terms of precision and recall. While our NLTK approach seems to have a slightly lower 
precision than EXEMPLAR  \cite{sonex}, it achieves a really higher recall, where the fraction of extracted relations relevant to the query is higher. This is the reason we preferred to
use our NLTK-based spatial relation method instead of EXEMPLAR.


\begin{table}[!htb]
\centering
\caption{Precision and recall for three different spatial relation extraction approaches.} \vspace{0.2cm}
\ra{1.1}
\begin{tabular}{l || c c c c}
\toprule
Method & \phantom{ab} & Precision & \phantom{ab} & Recall \\
\cmidrule{1-1} \cmidrule{3-3} \cmidrule{5-5}
EXEMPLAR \cite{sonex} & & $\textcolor[rgb]{0.4,0.1,0}{\mathbf{0.7}}$ & & $0.4$ \\
REVERB \cite{ReVerb} & & $0.1$ & & $0.4$ \\
NLTK  & & $0.6$ & & $\textcolor[rgb]{0.4,0.1,0}{\mathbf{0.82}}$ \\ 
\bottomrule
\end{tabular}\label{table:pre_recal}
\end{table}


Algorithm~\ref{alg:relext} describes the architecture of the proposed information extraction system. 
Initially, the raw text document is segmented into sentences  (Step 3). Each sentence is further subdivided (tokenized) into words and tagged as part-of-speech (Steps 5-6). 
Continuing name entities (POIs) are identified (Step 7). We will typically be looking for relations between specified types of named entities, 
which in NLTK are \emph{Organizations, Locations, Facilities and Geo-Political Entities (GPEs)}. 
In sequence, in case there are two or more name entities in the sentence, we check if any of the predefined syntactical patterns applies between the recognized name entity pairs (Step 12). 
If it applies, we then use regular expressions to determine the specific instance of the observed spatial relation from our predefined spatial relation pattern list for this case (Step 14).
If there is a string pattern match we record the extracted triplet (Steps 15-18). 
Thus, the search for spatial relations in texts results into a set of triplets $O$ of the form ($P_u$, $R_o$, $P_v$), where $P_u$ and $P_v$ 
are named entities of the required types and $R_o$ is the observed spatial relation that intervenes between $P_u$ and $P_v$. 


\IncMargin{1em}
\begin{algorithm}[!htb]
\DontPrintSemicolon
 \KwIn{A database of texts $T$, a set of syntactical patterns $A$, a set of spatial relation string patterns $R$}
 \KwOut{A set of triplets $O$ of the form $(P_{u}, ~R_o,~P_{v})$ where $P_{u} \neq P_{v}$ and $R_o \in R$}
 \BlankLine
 \Begin{

  \ForEach{\text{text} $t \in T$}{
   Extract sentences from $t$ into set $S$   

    \ForEach{sentence $s \in S$}{
      Token $s$ using NLTK \;
      PosTag $s$ using NLTK\;
      Identify name entities using NLTK

      \If{two or more name entities in $s$}{  
		Extract POI pairs in $P$
		
		\ForEach{$p \in P$}{
	   		$p_{A} \leftarrow$ Extract syntactical pattern of $p$
	    	
	    	\If{$p_{A} \in A$}{	
				$p_{R} \leftarrow$ Extract string pattern of $p$				

				\If{$ p_{R} \in R$}{
				$P_{u} \leftarrow p(1)$ \;
				$P_{v} \leftarrow p(2)$ \;
				$R_{o} \leftarrow p_{R}$ \;
		    		$O.\proc{PushTriplet}(P_{u},R_{o},P_{v})$
				}   
	      	}
	  	}
      }
    }
  }
  
\Return{$O$} \;
}
\caption{Spatial Relation Extraction}
\label{alg:relext}
\end{algorithm}
\DecMargin{1em}

\begin{figure}[h!] 
\centering
\includegraphics[width=0.65\textwidth,height=0.16\textwidth]{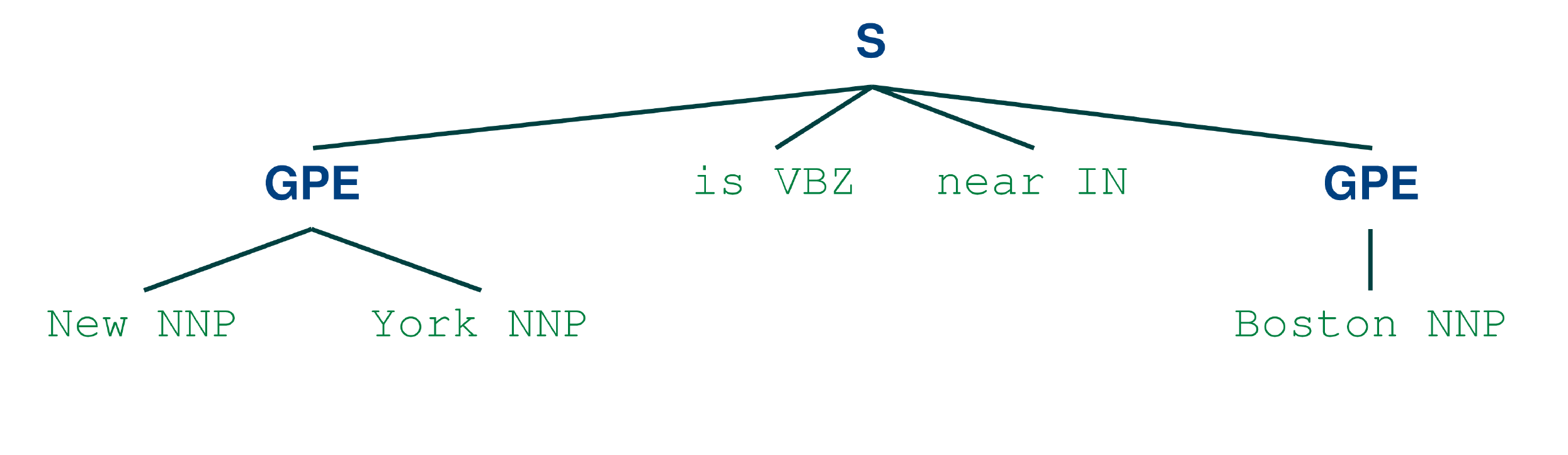} 
\caption{Example of a parsed sentence syntactic tree.}
\label{fig:parsetree}
\end{figure}
A relation extraction example is shown in Figure~\ref{fig:parsetree}, where the sentence is analyzed as explained and two named entities are identified as GPEs. 
We first check the syntax and make sure that the pattern 
\textit{``GPE - 3rd person verbal phrase (VBZ) - preposition/subordinating conjunction (IN) - GPE''} exists in our set of predefined spatial relation patterns. 
Performing string matching on the intermediate chunks (``near'') results in the triplet \textit{(New York, Near, Boston)}. 

\begin{figure}[t!]
 \centering
 \includegraphics[width=0.58 \textwidth]{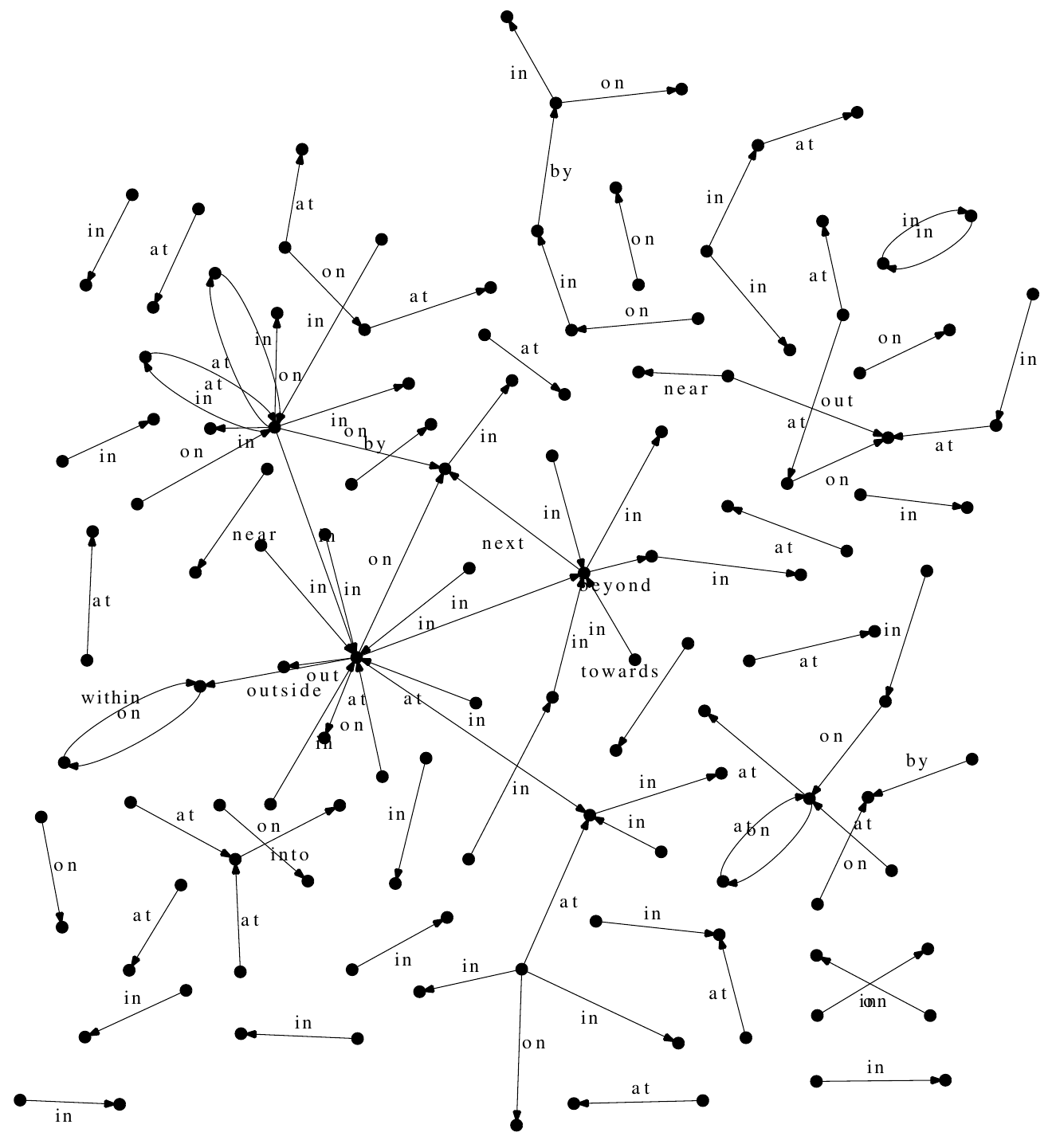}
 \caption{Small sample of a spatial relation graph for New York city.}
 \label{fig:relgraphs} 
\end{figure}

Applying Algorithm~\ref{alg:relext}, we extracted 440,000 triplets from our 120,000 travel blog text corpus. 
Figure~\ref{fig:relgraphs} shows a small sample of a \emph{Spatial Relationship Graph}, i.e., a spatial graph in which nodes represent POIs and edges label spatial relationships existing between them. 
The graph visualizes a sample of the spatial relationship data collected for New York city. In this work we extracted spatial relation data for four different cities, i.e. London, New York,
Paris and Beijing.
These four cases, going gradually from sparse to very dense spatial relation data, will be our main datasets during the experimental evaluation of the proposed approach. 
%


\subsection{Spatial Features}
\label{subsection:spatfeats}

Statistical models are often used to represent observations in terms of random variables. These models can then be used for estimation, description, and prediction based on basic probability theory. 
%
%
In our approach, we model a spatial relation between two POIs $P_u, P_v$
in terms of \emph{distance} and \emph{orientation}. We consider a labeled \emph{Spatial Feature Vector} as two random variables that model spatial relations in a probabilistic way. 

Assuming a projected (Cartesian) coordinate system, the distance is computed as the Euclidean metric between the two respective coordinates. 
The orientation is established as the counterclockwise rotation of the x-axis, centered at $P_v$, to point $P_u$. 

Several instances of a spatial relation are used to create a dataset, which will be used to train one probabilistic  
model for each spatial relation. 
For a concise and consistent mathematical formalization, let us consider that for each instance of each relation, we create a two-dimensional spatial feature vector $X = (X_d, X_o)^{\intercal}$ where $X_d$ denotes 
the distance and $X_o$ denotes the orientation between $P_u$ and $P_v$.
%
This way, we end up with a set of two-dimensional feature vectors $\mathcal{X}=\{X_1, X_2, \dots, X_n\}$ for each spatial relation.

\section{Spatial Relationship Modeling} 
\label{sec:modeling}

In this section we describe the probabilistic modeling we follow in order to quantify qualitative spatial data. 
Key ingredients of our system are methods that train probabilistic models. 
Our analysis below includes 
$(i)$ a short description of the probabilistic mixture models we employ for the quantitative representation of spatial relations 
(Section \ref{subsection:gmmem}) and, $(ii)$ a greedy learning algorithm for model parameter estimation (Section \ref{subsection:modopt}). 
Overall, this section describes a method that trains probabilistic models that quantify crowdsourced spatial relationships. 
These estimates can then be used to reason about unknown POI locations in textual narratives (see Section \ref{sec:evaluation}).

\vspace{5pt}
\subsection{Quantifying Qualitative Relations}
\label{subsection:gmmem}
An essential step in quantifying qualitative data is the mapping of the generated data to pre-selected probability density functions (PDFs).
In \cite{LiBarronMixDensity}, 
it is shown that for any heterogeneous mutli-dimensional data that originates from an \emph{arbitrary} PDF $p$, there exists a sequence 
of finite mixtures $p_k(x) = \sum_{i=1}^{k} w_{i} g(x;\theta_{i})$ that
achieves Kullback-Leibler (KL) divergence $$D(p||p_{k}) - D(p||g_{p}) \leq  \mathcal{O}(1/k)$$ for any $g_p = \int g(x;\theta) P(d\theta)$.
i.e., one can achieve a good approximation of $p$ with rate $\mathcal{O}(1/k)$ by using a $k$-component mixture of $g(x;\cdot)$.  
Furthermore, this bound is achievable by employing a greedy training scheme \cite{LiBarronMixDensity}, i.e., we can approximate any density $p$ by a greedy training procedure. 

In this work we employ Gaussian Mixture Models (GMMs) which have been extensively used in many classification and general machine learning problems (cf. \cite{Bishop}). 
They are very well known for $(i)$ their formality, as they build on the formal probability theory, $(ii)$ their practicality, as they have been implemented several times in practice, 
$(iii)$ their generality, as they are capable of handling many different types of uncertainty, and $(iv)$ their effectiveness. 
 
In general, a GMM is a weighted sum of $M$ component Gaussian densities as $p(x|\lambda) = \sum_{i=1}^{M} w_{i} g(x;\mu_{i},\Sigma_{i})$
where $x$ is a $d$-dimensional data vector (in our case $d=2$), $w_i$ are the mixture weights, and $g(x;\mu_i, \Sigma_i )$ is a Gaussian density function 
with mean vector $\mu_i \in \mathbb{R}^{d} $ and covariance matrix $\Sigma_i \in \mathbb{R}^{d\times d}$. 
To fully characterize $f$, one requires the mean vectors, the covariance matrices and the mixture weights. These parameters are collectively represented 
in $\lambda = \{w_i, \mu_i, \Sigma_i\}$ for $i= 1, \dots, M$. 

In our setting, each spatial relation is modeled by a 2-dimensional GMM, trained on each relation's spatial feature vectors, 
as detailed in Section~\ref{subsection:spatfeats}.
For the parameter estimation of each Gaussian component of each GMM, we use Expectation Maximization (EM) (cf. \cite{Dempster}). 
EM enables us to update the parameters of a given M-component mixture with respect to a feature vector 
set $\mathcal{X} = \{X_1 , \dots, X_n \}$ with $1 \leq j \leq n$ and all $X_j \in \mathbb{R}^{d}$, such that the log-likelihood 
$\mathcal{L} = \sum_{j=1}^{n} \log(p(X_j|\lambda))$ 
increases with each re-estimation step, i.e., EM re-estimates model parameters $\lambda$ until $\mathcal{L}$ convergence. 
 

%
%
%
%


\vspace{8pt}
\subsection{Model Optimization}
\label{subsection:modopt}
\vspace{5pt}

A main issue in probabilistic modeling with probability mixtures is that a predefined number of components is neither a dynamic nor an efficient and robust approach. 
The optimal number of components should be decided based on each dataset.  

We employ a greedy learning approach to dynamically estimate the number of components in a GMM. (cf. \cite{Verbeek}). 
Our approach builds the mixture component in an efficient way by starting from an one-component GMM---whose parameters are trivially computed by using EM--- and 
then employing the following two basic steps until a stopping criterion is met:

\begin{enumerate}
\item Insert a new component in the mixture 
\item Apply EM until the log-likelihood $\mathcal{L}$ or the parameters of the GMM converge (cf. Section~\ref{subsection:gmmem})
\end{enumerate}

The stopping criterion can either be a maximum pre-selected number of components, or it can be any other model selection criterion. 
%
In our case the algorithm stops if the maximum number of components is reached, 
or if the new model's log-likelihood $\mathcal{L}+1$ is less or equal to the log-likelihood $\mathcal{L}$ of the previous model, 
after introducing a new component. 

For a more formal description let us consider a feature vector set $\mathcal{X}$ under a $M$-component mixture $p^{M}(\mathcal{X}|\lambda)$. 
The greedy learning procedure can be summarized in Algorithm~\ref{alg:gmmopt}. For each spatial feature vector, we estimate the parameters and the 
log-likelihood of an one-component model (Steps 4-5). In sequence, we find a new component and add it to the previous mixture (Steps 7-8). 
Then, we re-estimate the model parameters and log-likelihood (Steps 9-10) until we reach the desiderata described above. 

\IncMargin{1em}
\begin{algorithm}[h!]
 \DontPrintSemicolon
 \KwIn{A set of spatial feature vectors $\mathcal{\hat{X}}$, a maximum number of components in $\mathcal{M_{C}}$} 
 \KwOut{A set of trained GMMs $\mathcal{\hat{G}}$} 
 
 \BlankLine

 \Begin{
  
  $M \leftarrow 1$\;
  
  \ForEach{$\mathcal{X} \in \mathcal{\hat{X}}$}{
      $p^{M}(\mathcal{X}|\lambda) \leftarrow$ Estimate 1-component model parameters using EM\;
      $\mathcal{L}^{M} \leftarrow$ Calculate 1-component model log-likelihood\;
   \While{$M \leq \mathcal{M_{C}}$}{
	$g(\mathcal{X};\lambda^{\ast}) \leftarrow$ Optimal new component for $(p^{M}(\mathcal{X}|\lambda))$\;
	$p^{M+1}(\mathcal{X}|\lambda) \leftarrow$ Combine model  $p^{M}(\mathcal{X}|\lambda)$ and component $g(\mathcal{X};\lambda^{\ast})$ in a new model\;
	$p^{M+1}(\mathcal{X}|\lambda) \leftarrow$ Estimate new model parameters using EM\;
        $\mathcal{L}^{M+1} \leftarrow$ Calculate new model log-likelihood\;
        
        \eIf{$\mathcal{L}^{M+1} \leq \mathcal{L}^{M}$}{
	    $\mathcal{\hat{G}}.\proc{PushGMM}(p^{M}(\mathcal{X}|\lambda))$\;
	    $\proc{Terminate}()$\;
	    }{
	     $M \leftarrow M +1$\;	    
	    }

      }
      $\mathcal{\hat{G}}.\proc{PushGMM}(p^{M}(\mathcal{X}|\lambda))$\;
    }
  
\Return{$\mathcal{\hat{G}}$} \;
}
\caption{Optimized GMM Training}
\label{alg:gmmopt}
\end{algorithm}
\DecMargin{1em}

The crucial step of this algorithm is the search for an optimal new component (Step 7). 
Several approaches exist for this issue: One is to consider a number of candidates equal to the number of 
feature vectors but it is identified that such strategy would be rather expensive. 
The approach followed in this work is to pick an optimal number of candidate components as discussed in \cite{Verbeek}.  
More specifically, for each insertion problem in a $k$-component mixture, the dataset $\hat{\mathcal{X}}$ is partitioned in $k$ disjoint subsets
and a fixed number $m$ of candidate components is generated per existing mixture component, e.g., for a $k$-component mixture $k \times m$ 
candidate components are generated. In our experiments we used $m=10$. Finally, with the use of EM algorithm we pick the candidate component
that maximizes the log-likelihood $\mathcal{L}+1$ when mixed into the previous mixture $p^{M}(\mathcal{X}|\lambda)$.

\section{Location Fusion} 
\label{sec:evaluation}

Based on the discussion above, we introduce a location prediction system that employs the proposed spatial relation modeling approach
described in Section~\ref{sec:modeling}. All the text processing parts are implemented in Python while modeling and experimentation parts 
where implemented in Matlab. Our tests were conducted on an Intel(R) Core(TM) i5-2400 CPU at 3.10GHz with 8GB of RAM, running Ubuntu Linux 12.10.

\subsection{Quantitative and Qualitative Performance} \label{subsec:randompoint} 
Our goal with this experiment is to show that our approach performs well for various location estimation scenarios using crowdsourced data. In particular, we assume that unknown locations in relation to known POIs do exist in crowdsourced data. 
The first scenario, e.g. random point location prediction, uses randomly generated POIs and tries to estimate their locations by fusing spatial relationship observations to known POIs (landmarks). 

The procedure shown in the pseudocode of Algorithm~\ref{alg:randpopre} is as follows.
%
%
At first, we partition the space with respect to grid vertices (landmarks) (Step 2). For example, for a grid dimensionality $\mathcal{G_{D}}$ = \emph{15}, we have \emph{15$\times$15 = 225} grid vertices and \emph{14$\times$14 = 196} grid cells (regions). 
We generate a random point $\mathcal{R_{P}}$ (Step 5) and, for each vertex (landmark) $gv$, starting with the bottom-left vertex and proceeding then to the right and in a row-by-row fashion, 
we find the spatial relation model $\mathcal{\hat{G}}^{'}$ 
that maximizes the likelihood---in terms of probability---of $\mathcal{R_{P}}$ (Step 8).
Then, the selected spatial relation model $\mathcal{\hat{G}}^{'}$ is used to assign positional probabilities to each other vertex (landmark) $gv'$ of the grid using the current vertex as a reference point (Steps 10-13). In this way, 
we update the likelihoods of all vertices in the grid.
All these likelihoods are stored in matrix $\mathcal{V_{L}}$ (Step 11). Finally, all the likelihoods per vertex are summed up and normalized in $\mathcal{F_{VL}}$ (Step 16) and a probability is assigned 
to each grid cell (region) $\mathcal{R_{L}}$ (Step 17). The overall likelihood of each grid cell is calculated as the mean value of the likelihoods of its four vertices. In a final step, 
we keep track of how many times the region that contains the unknown point is ranked among the $K$-highest (Top-K) probable regions (Steps 18-20). 
This allows us to measure the prediction accuracy of the proposed approach (Step 22).

\vspace{6pt}
\IncMargin{1em} 
\begin{algorithm}[!th] 
\DontPrintSemicolon \KwIn{A set of trained GMMs $\mathcal{\hat{G}}$, a bounding box $\mathcal{B_{B}}$, grid dimensionality in $\mathcal{G_{D}}$, a number $\mathcal{R_{N}}$ of random points to be generated} 
\KwOut{Prediction accuracy $\mathcal{A}$} 
\BlankLine
	
\Begin{



	
$\mathcal{G_{V}} \leftarrow$ Calculate grid vertices for $\mathcal{B_{B}}$ based on $\mathcal{G_{D}}$\;
	
$\id{InTop5} \leftarrow 0$\;

	
\For{$i\leftarrow 1$ \KwTo $\mathcal{R_{N}}$}{
	
$\mathcal{R_{P}} \leftarrow$ Generate a random point in $\mathcal{B_{B}}$\; $\id{Indx1} \leftarrow 0$\;
	
\ForEach{$gv \in \mathcal{G_{V}}$}{ \BlankLine
	
$\mathcal{\hat{G}}^{'} \leftarrow \arg\max\limits_{g\in\mathcal{\hat{G}}} P(\mathcal{R_{P}}|g, gv)$\; $\id{Indx2} \leftarrow 0$\;
	
\ForEach{$gv^{'} \in \mathcal{G_{V}}$}{ 
$\mathcal{V_{L}}(\id{Indx1,Indx2}) \leftarrow$ Calculate each $gv'$ vertex's likelihood with $gv$ as reference given model $\mathcal{\hat{G}}^{'}$\; 
$\id{Indx2} \leftarrow \id{Indx2} + 1$\; } $\id{Indx1} \leftarrow \id{Indx1} + 1$\; } 
$\mathcal{F_{VL}} \leftarrow$ Sum and normalize each vertex's likelihoods in $\mathcal{V_{L}}$\; 
$\mathcal{R_{L}} \leftarrow$ Calculate each region's likelihood using $\mathcal{F_{VL}}$\;
	
	
\If{$\mathcal{R_{P}}$ in $K$ highest probability $\mathcal{R_{L}}$}{
	
$\id{InTopK} \leftarrow \id{InTopK} + 1$\; } }
	
$\mathcal{A} \leftarrow \proc{Percent}(InTopK)$\;


\Return{$\mathcal{A}$}\;} 
\caption{Random Point Location Prediction} \label{alg:randpopre} 
\end{algorithm}
\DecMargin{1em}

\begin{figure*}[!th] 
\subfigure[]{
\includegraphics[width=0.349
\textwidth,height=1.4in]{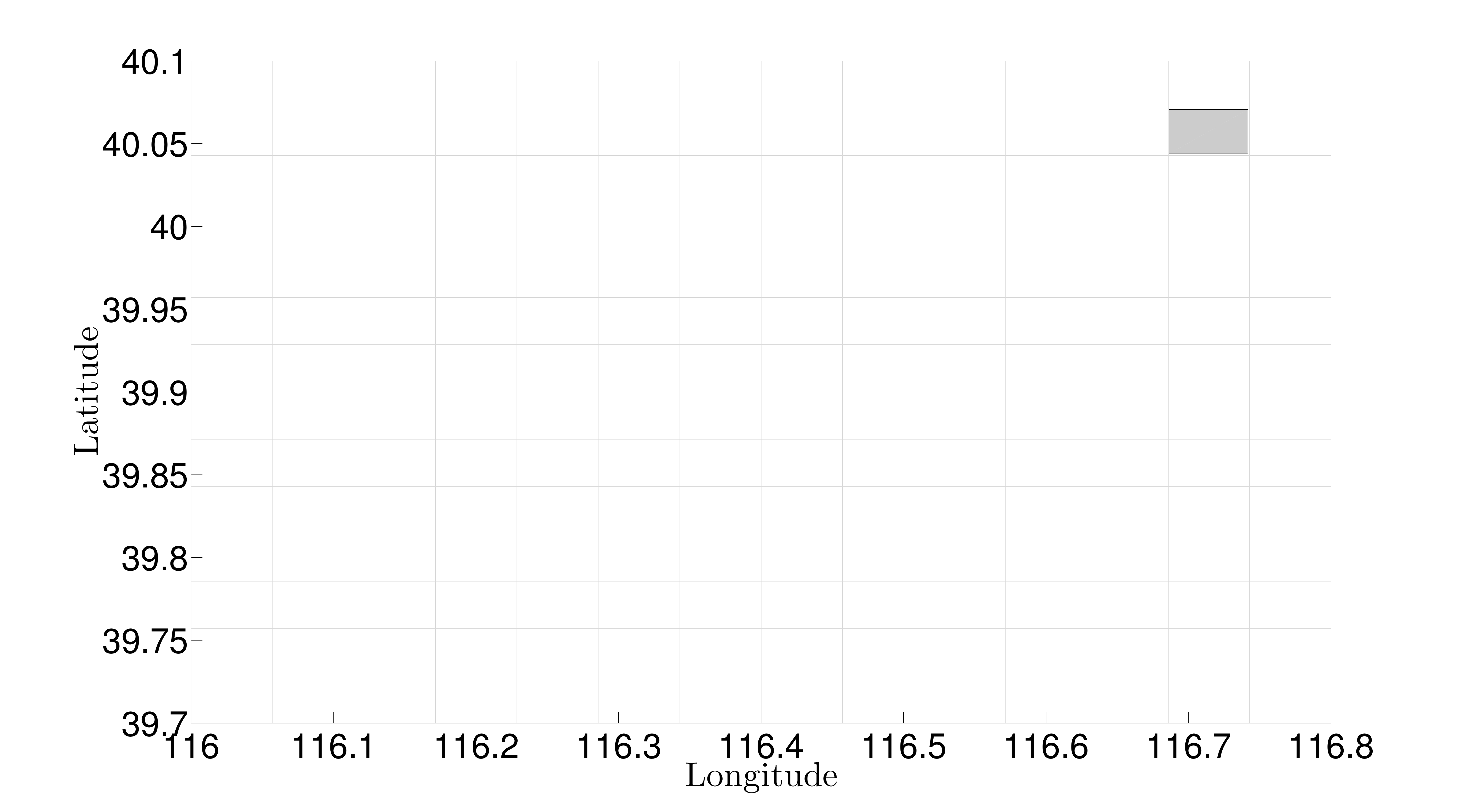}\label{subfig:beijingunkn}} \hspace{-0.5cm} \subfigure[]{
\includegraphics[width=0.349
\textwidth,height=1.4in]{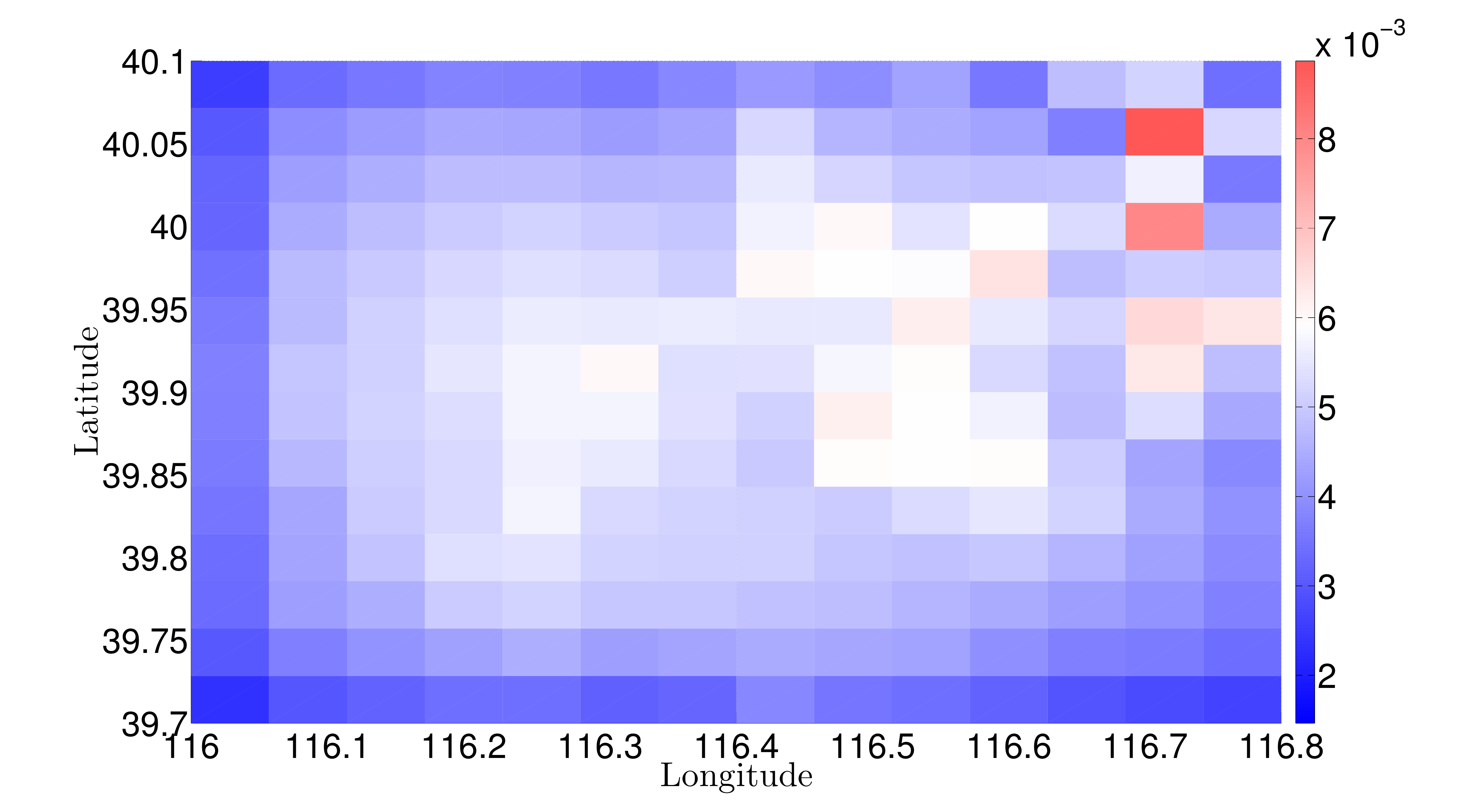}\label{subfig:beijingpred}} \hspace{-0.35cm} \subfigure[]{
\includegraphics[width=0.349
\textwidth,height=1.4in]{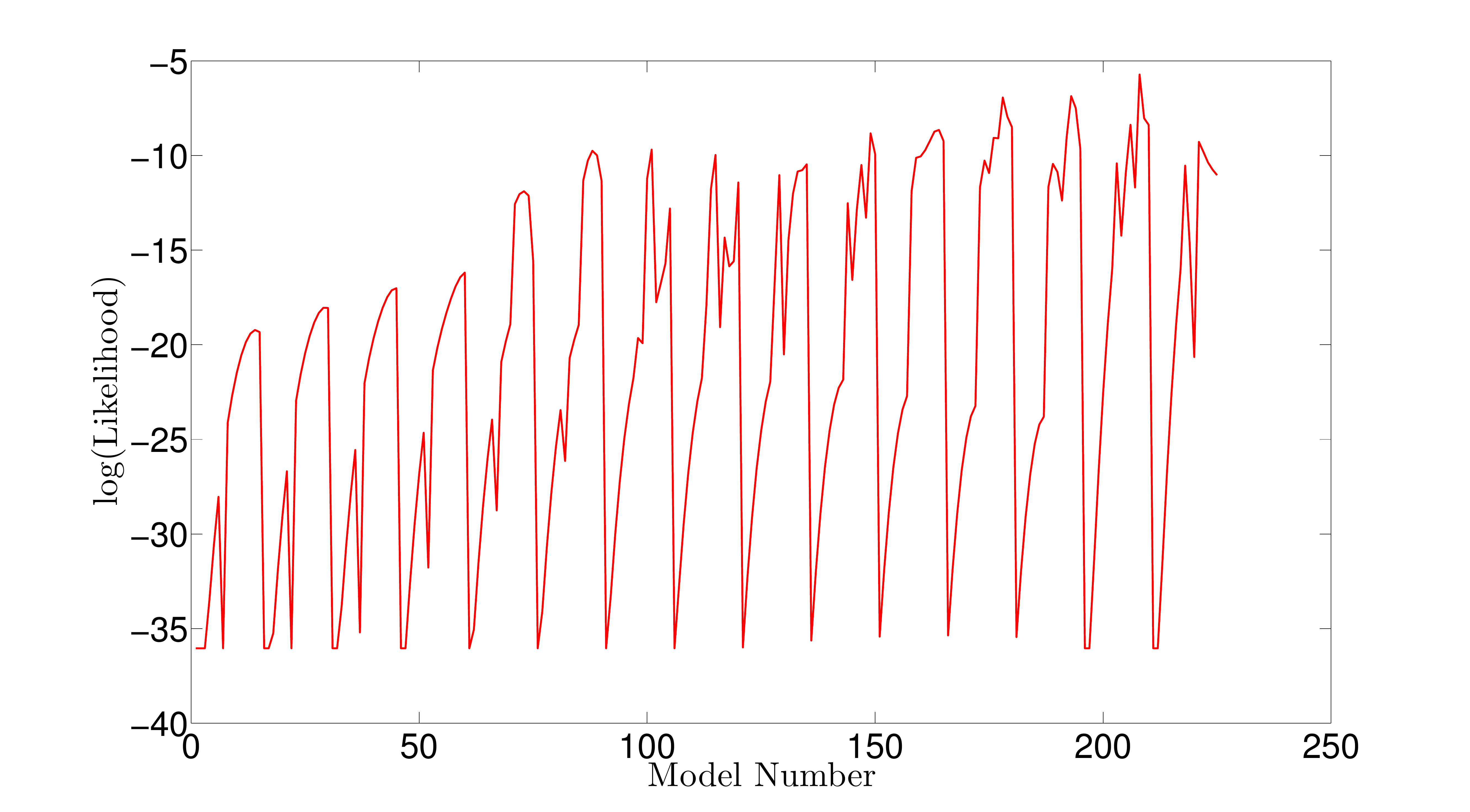}\label{subfig:beijinglogprobmonitor}}\\
\subfigure[]{
\includegraphics[width=0.498
\textwidth]{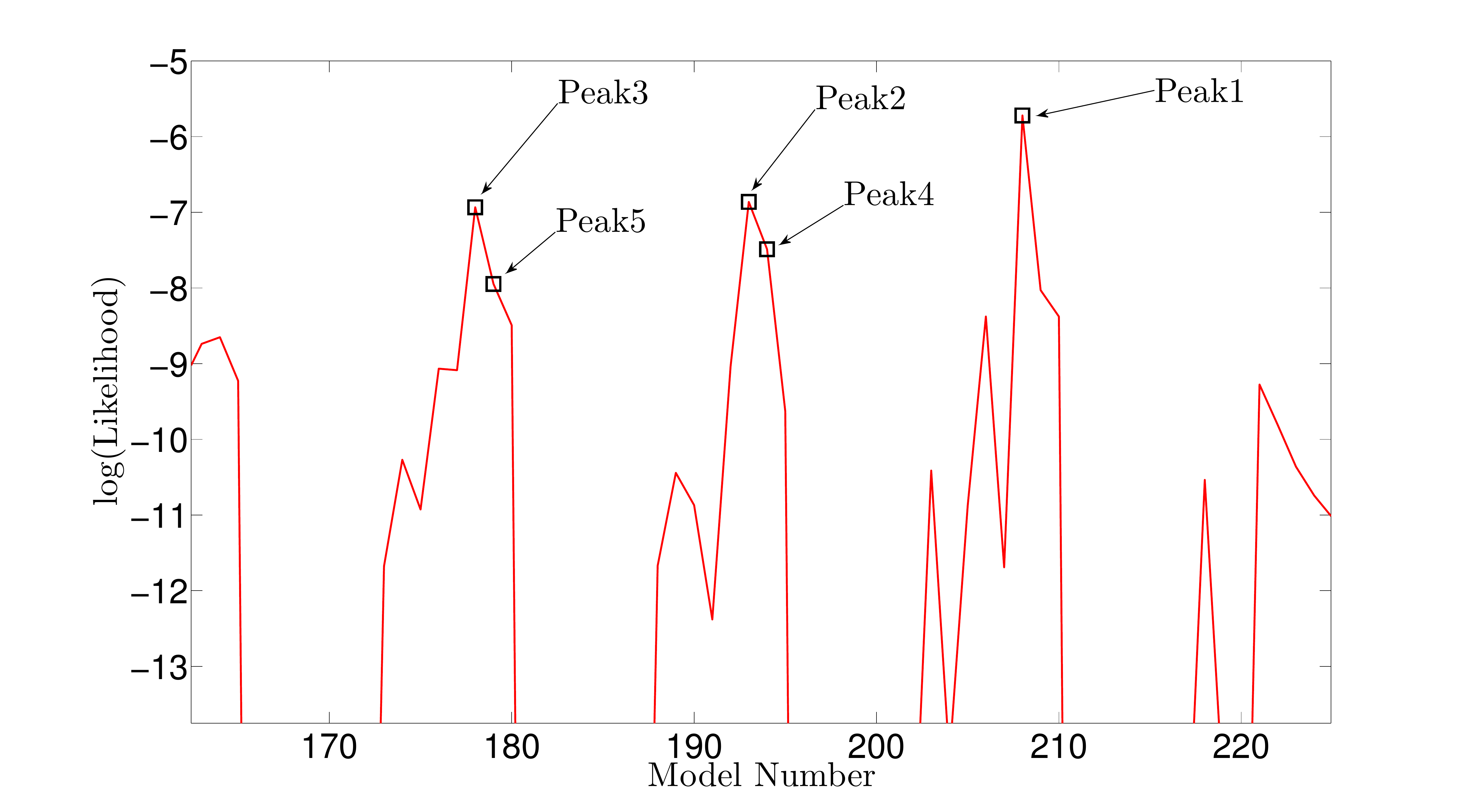}\label{subfig:zoom1}} \subfigure[]{
\includegraphics[width=0.498
\textwidth]{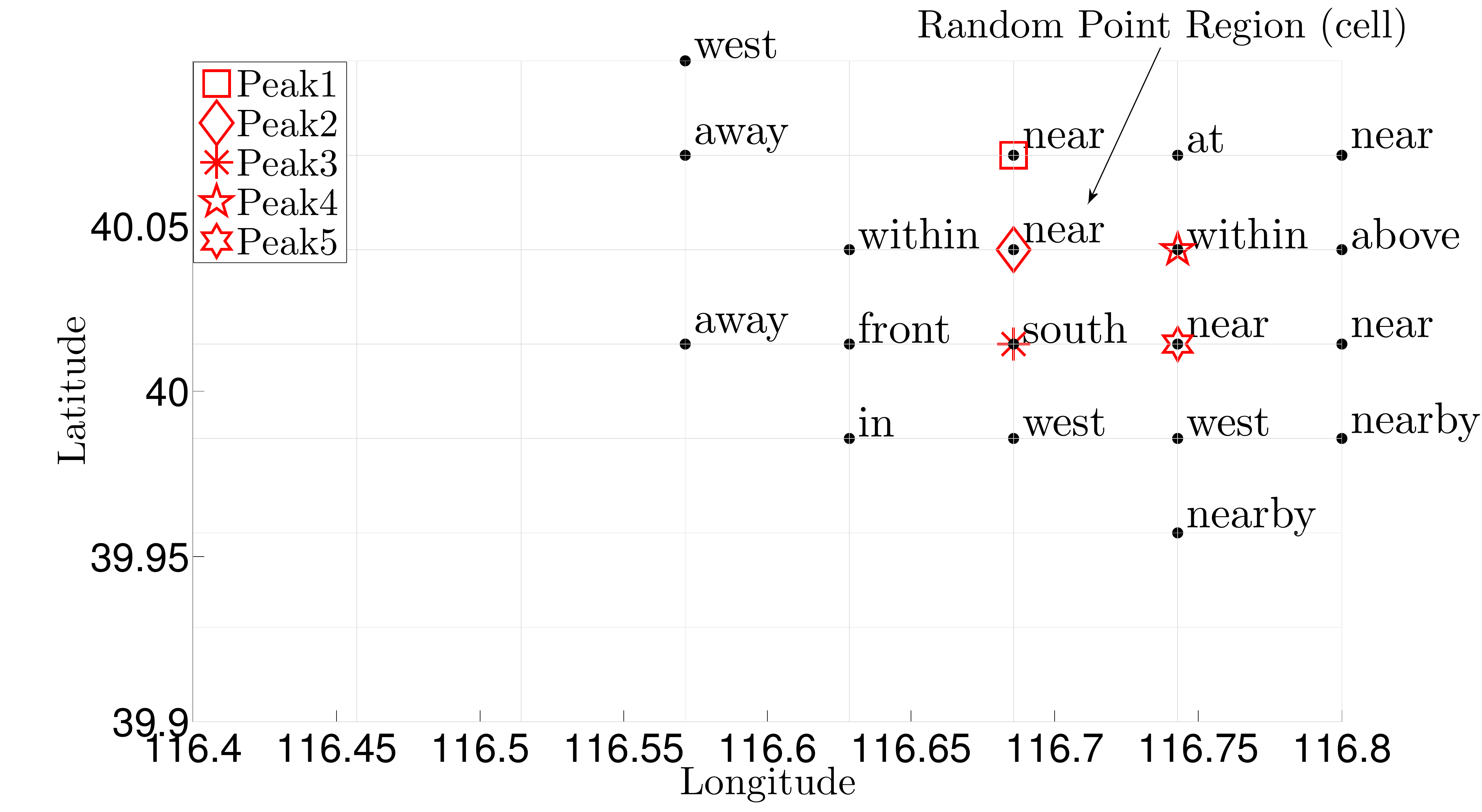}\label{subfig:zoom2}} 
\caption{A location prediction scenario: (a) illustrates the region in which a random point has been generated, 
(b) visualizes using heatmap colors the probability of each region after a full run of Algorithm~\ref{alg:randpopre}, 
(c) illustrates the log-Likelihood of the random point's region as we traverse from one vertex (landmark) to the other, 
(d) shows an enlarged portion of (c) with the five highest likelihood peaks, and (e) illustrates the 20 best vertex-models and emphasizes the 5 best vertex-models (peaks in (d)).\vspace{5pt}} \label{fig:quantitative} 
\end{figure*}

Figure~\ref{fig:quantitative} provides a more elaborate description of a simulation for very challenging case of Beijing. 
The light gray colored grid cell in Figure~\ref{subfig:beijingunkn} illustrates the region in which a random point was generated. 
Figure~\ref{subfig:beijingpred} shows the assigned probabilities to each region after a full run of Algorithm~\ref{alg:randpopre}. 
We observe that our approach assigns the highest probability for the random point location by using only spatial relation information extracted with respect to the landmarks in the region. 

Figure \ref{subfig:beijinglogprobmonitor} illustrates the monitoring of the log-likelihood of the random point region as we sequentially visit each vertex (landmark) in the grid. 
Starting from the lower left grid vertex ($\mathcal{\hat{G}}^{'}$ model 1) and proceeding row-wise until the upper right vertex ($\mathcal{\hat{G}}^{'}$ model 225), 
the log-likelihood increases as we move closer to the desired region and decreases when moving away. Figure~\ref{subfig:zoom1} points out the five highest 
likelihoods of the random point region and Figure~\ref{subfig:zoom2} the locations of these vertices (landmarks) in the grid along with the
model's textual description.
%
This analysis should be considered a qualitative accuracy assessment. Most of the models selected are qualitatively correct and express a real spatial relation between each 
corresponding vertex and random point. The relations shown in Figure~\ref{subfig:zoom2} are of the form $(P_u, R_o, P_v)$ as 
presented in Section~\ref{subsec:data2}. 

\begin{figure*}[!th] 
\subfigure[]{ \includegraphics[width=0.52 \textwidth]{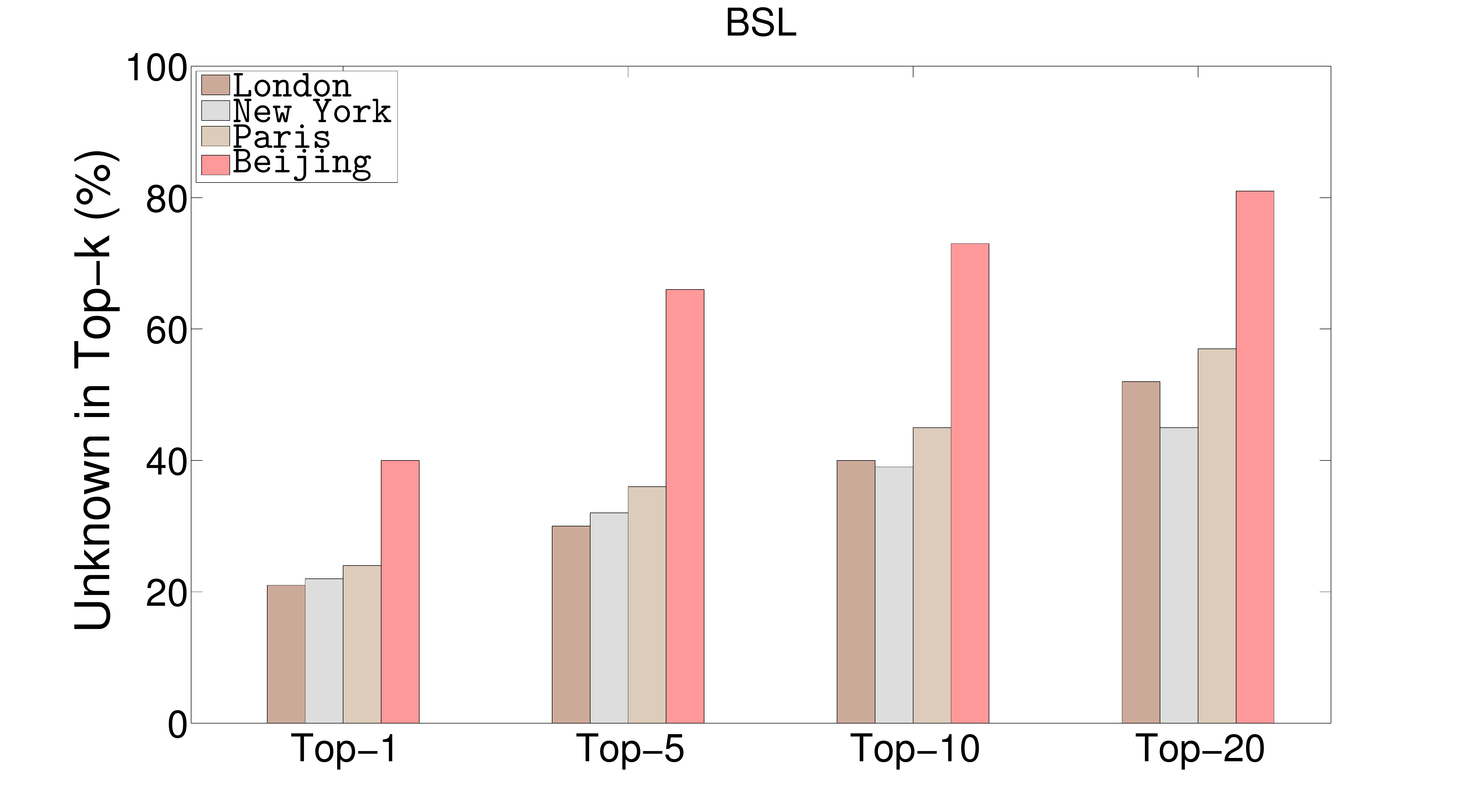} \label{subfig:pa_bsl}}\hspace{-0.3cm} 
\subfigure[]{\includegraphics[width=0.52 \textwidth]{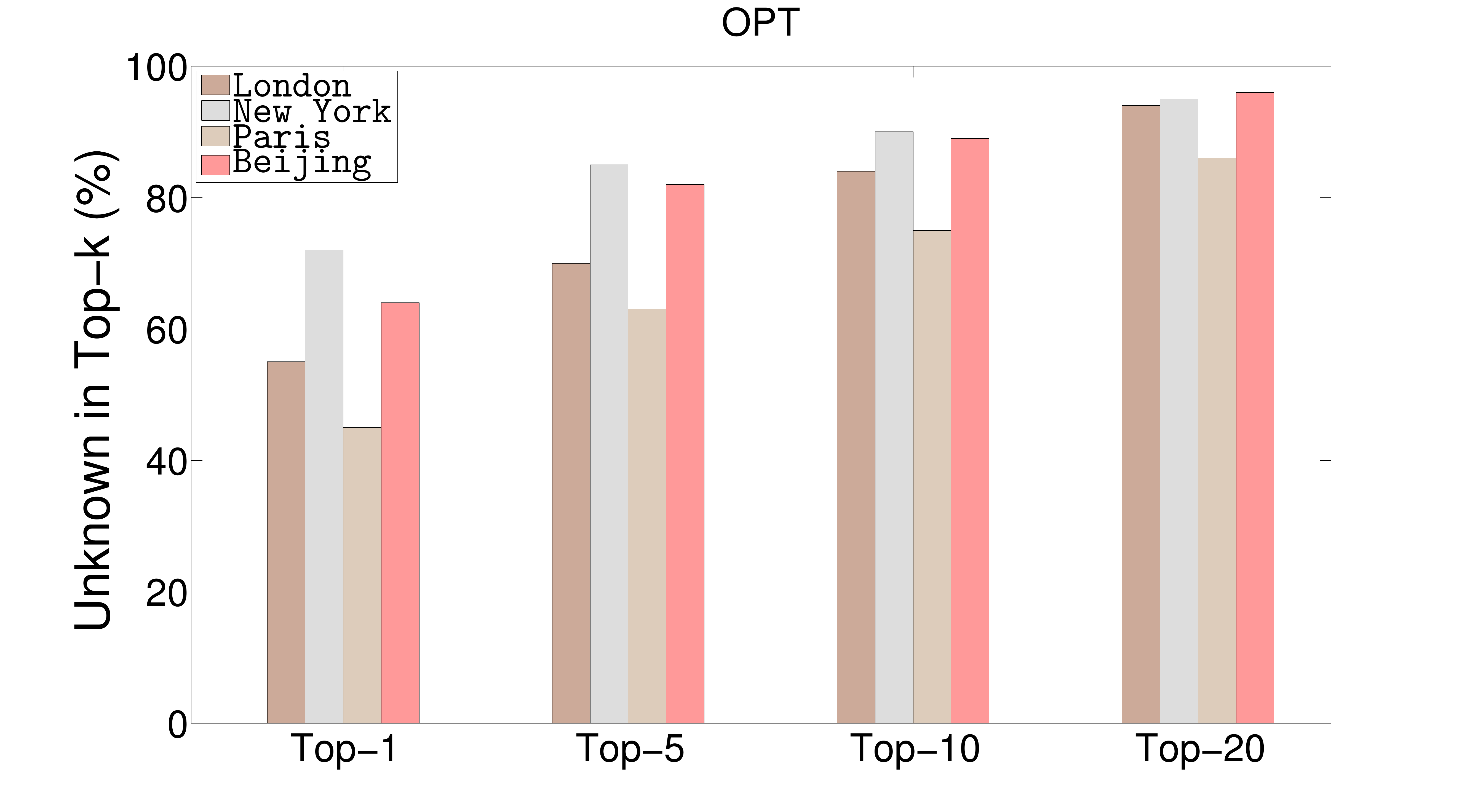} \label{subfig:pa_opt}}
\caption{Location prediction accuracy. (a) Illustrates the prediction accuracy of the BSL model for $K$ values $1, 5, 10, 20$ respectively.
(b) Illustrates the prediction accuracy of the OPT model for $K$ values $1, 5, 10, 20$ respectively. \vspace{10pt}} 
\label{fig:quant} 
\end{figure*}

To the best of our knowledge, this is the first work on location prediction based on textual descriptions. Consequently, in order to provide some comparison results, we define a $1-component$ \textbf{baseline} (BSL) model 
(a GMM model $p(x|\lambda) = \sum_{i=1}^{M} w_{i} g(x;\mu_{i},\Sigma_{i})$ with $M=1$), and an optimized model (OPT) trained as analyzed in Section~\ref{subsection:modopt} and Algorithm~\ref{alg:gmmopt}.  
We run Algorithm~\ref{alg:randpopre} for both BSL and OPT models, for all four datasets, with 1000 random points per dataset. 
Additionally, we consider the cases where the randomly generated point's region is among the $TopK$ predicted regions with 
$K$ values $1,5,10,20$ respectively. The prediction accuracy results are shown in Figure~\ref{fig:quant}. Figure~\ref{subfig:pa_bsl}
illustrates the prediction accuracy of the BSL model while Figure~\ref{subfig:pa_opt} illustrates the prediction accuracy
of the OPT model. The results show the superiority of our model, as opposed to the 1-component model. 
Additionally, Table~\ref{table:quant_gain}
shows the actual prediction accuracy improvement when we use the OPT model. In some cases (indicated in bold) 
the prediction accuracy improvement is equal to or greater than $30\%$. 
These results show that ``colloquial'' location estimation facilitated by crowdsourced geospatial narratives is a feasible approach. 

\begin{table}[!h]
\centering
\caption{Prediction accuracy improvement when the optimized model (OPT) is used instead of the BSL model.} \vspace{0.2cm}
\ra{1.1}
\begin{tabular}{l || c c c c c c c c}
\toprule
& \multicolumn{8}{c}{Improvement per Top-k case} \\
\toprule
Dataset && $k=1$ &&  $k=5$ && $k=10$ &&  $k=20$ \\

\cmidrule{1-1} \cmidrule{3-3} \cmidrule{5-5} \cmidrule{7-7} \cmidrule{9-9}
London   &&  $+\textcolor[rgb]{0.4,0.1,0}{\mathbf{34}}\%$	   && $+\textcolor[rgb]{0.4,0.1,0}{\mathbf{40}}\%$  &&  $+16\%$	&& $+15\%$\\
New York &&  $+\textcolor[rgb]{0.4,0.1,0}{\mathbf{50}}\%$       && $+\textcolor[rgb]{0.4,0.1,0}{\mathbf{53}}\%$  &&  $+\textcolor[rgb]{0.4,0.1,0}{\mathbf{51}}\%$	&& $+\textcolor[rgb]{0.4,0.1,0}{\mathbf{50}}\%$\\
Paris    &&  $+21\%$       && $+27\%$  &&  $+\textcolor[rgb]{0.4,0.1,0}{\mathbf{30}}\%$	&& $+29\%$\\ 
Beijing  &&  $+24\%$       && $+16\%$  &&  $+16\%$	&& $+15\%$\\ 
\bottomrule
\end{tabular}\label{table:quant_gain}
\end{table}
\vspace{10pt}

\begin{figure}[!th] \centering
\includegraphics[width=0.5\textwidth,height=1.8in]{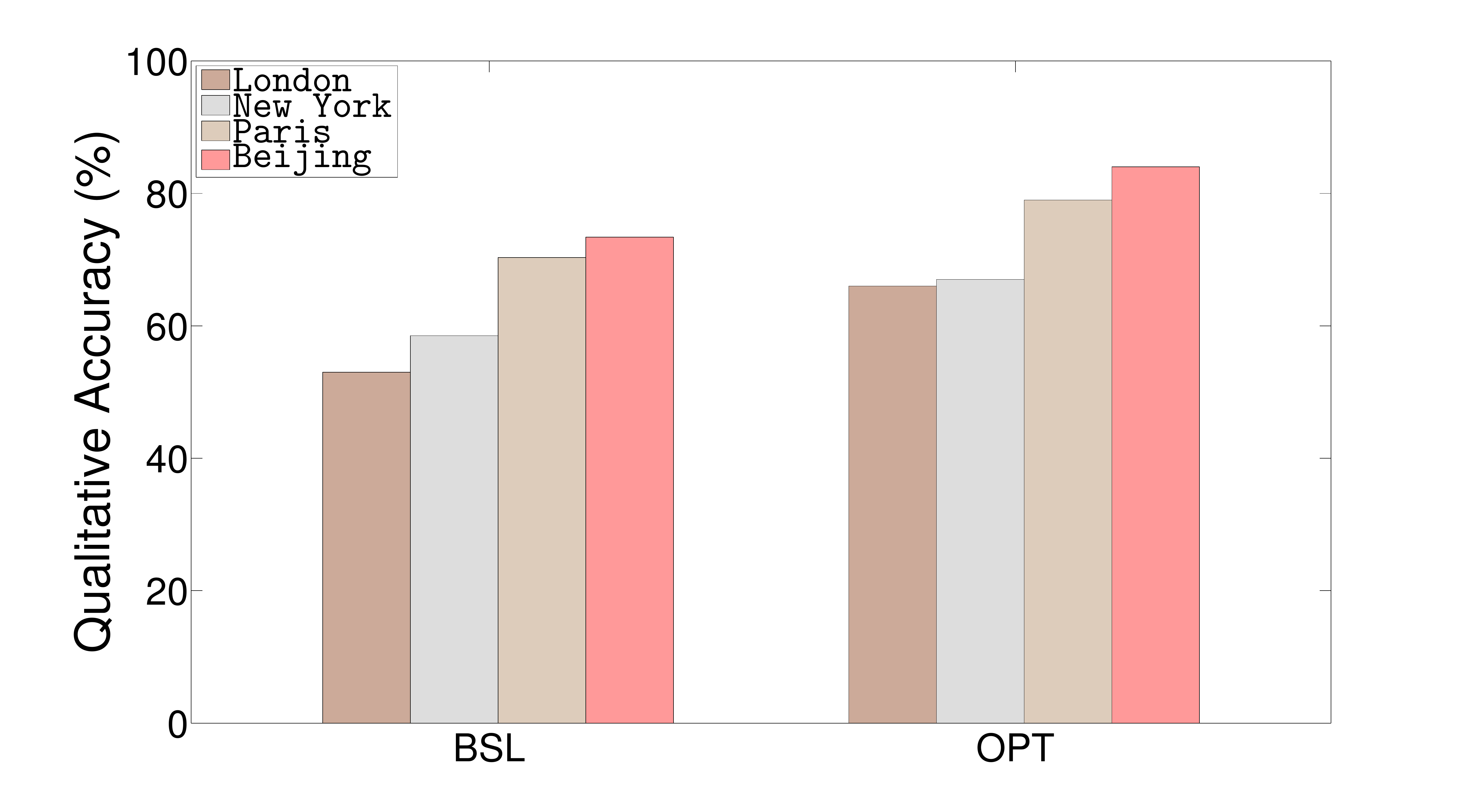} 
\caption{Illustrates the percentage of qualitatively correct spatial relations for the BSL and OPT models respectively.} 
\label{fig:qual} 
\end{figure}

Finally, we also want to measure the percentage of selected models $\mathcal{\hat{G}}^{'}$ that are qualitatively correct, i.e., they reveal a true spatial relation between a vertex and a random point. 
%
%
Figure~\ref{fig:qual} shows the percentage of the selected models $\mathcal{\hat{G}}^{'}$ that depict an accurate spatial relation between the vertices and random points for
both BSL and OPT models. As in the prediction accuracy case, the qualitative accuracy of the OPT model is quite higher than that of the BSL model. 
Table~\ref{table:qual_acc} shows this improvement of OPT model over BSL model in relative terms. In some cases (indicated in bold)
the qualitative accuracy improvement is more than $10\%$.   
%

\begin{table}[!h]
\centering
\caption{Qualitative accuracy improvement when the optimized model (OPT) is used instead of the BSL model.} \vspace{0.2cm}
\ra{1.1}
\begin{tabular}{l || c c c c}
\toprule
Dataset & & \phantom{ab} & & Improvement \\
\cmidrule{1-1} \cmidrule{5-5} 
London    & & & & $+8\%$ \\
New York  & & & & $+\textcolor[rgb]{0.4,0.1,0}{\mathbf{11}}\%$ \\
Paris     & & & & $+\textcolor[rgb]{0.4,0.1,0}{\mathbf{13}}\%$ \\ 
Beijing   & & & & $+9\%$ \\ 
\bottomrule
\end{tabular}\label{table:qual_acc}
\end{table}

\begin{figure*}[!th] 
\subfigure[\textbf{Near} the grid center]{
\includegraphics[width=0.34
\textwidth,height=1.4in]{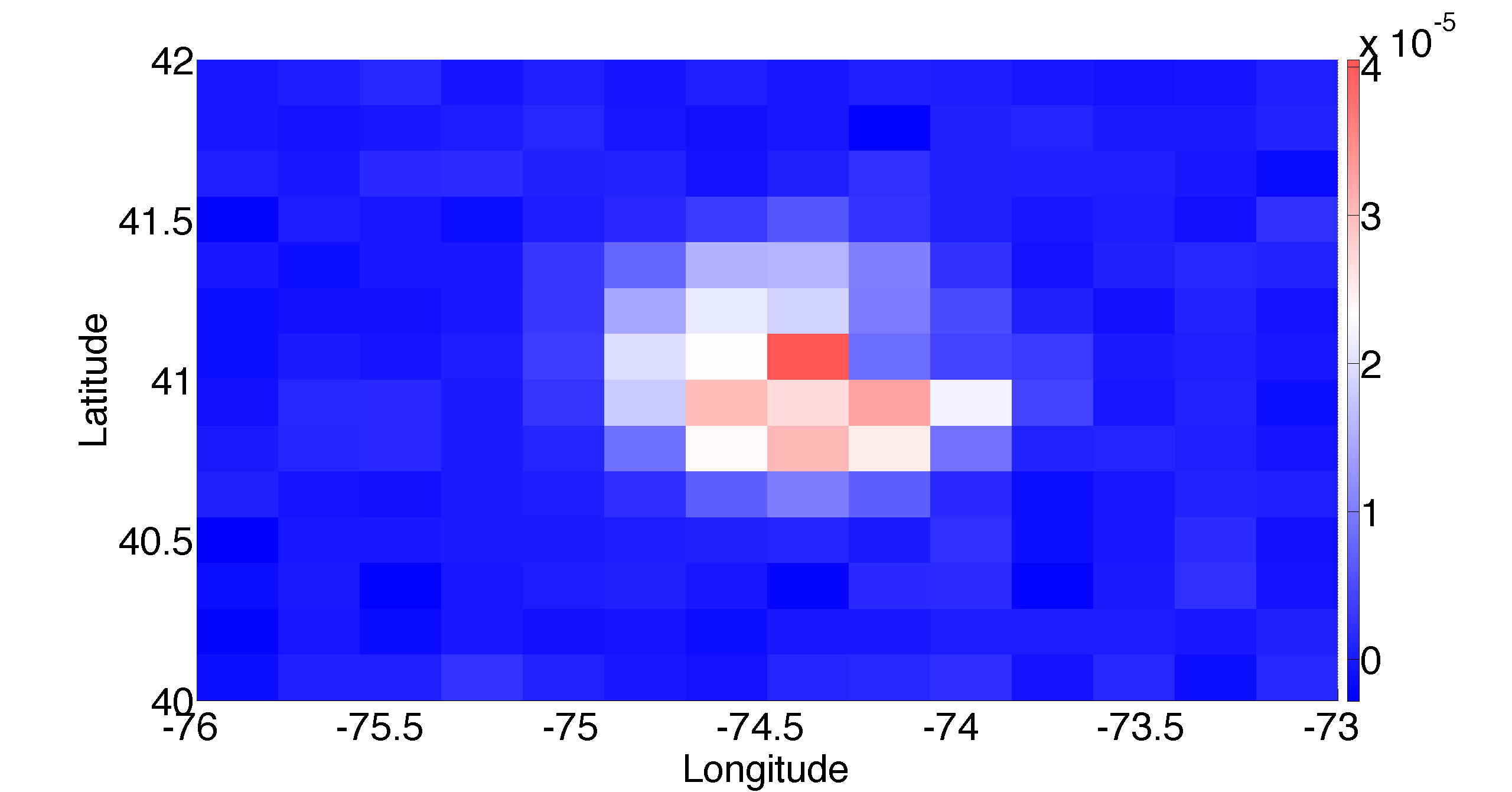}} \hspace{-0.3cm} \subfigure[\textbf{At} the grid center]{
\includegraphics[width=0.34
\textwidth,height=1.4in]{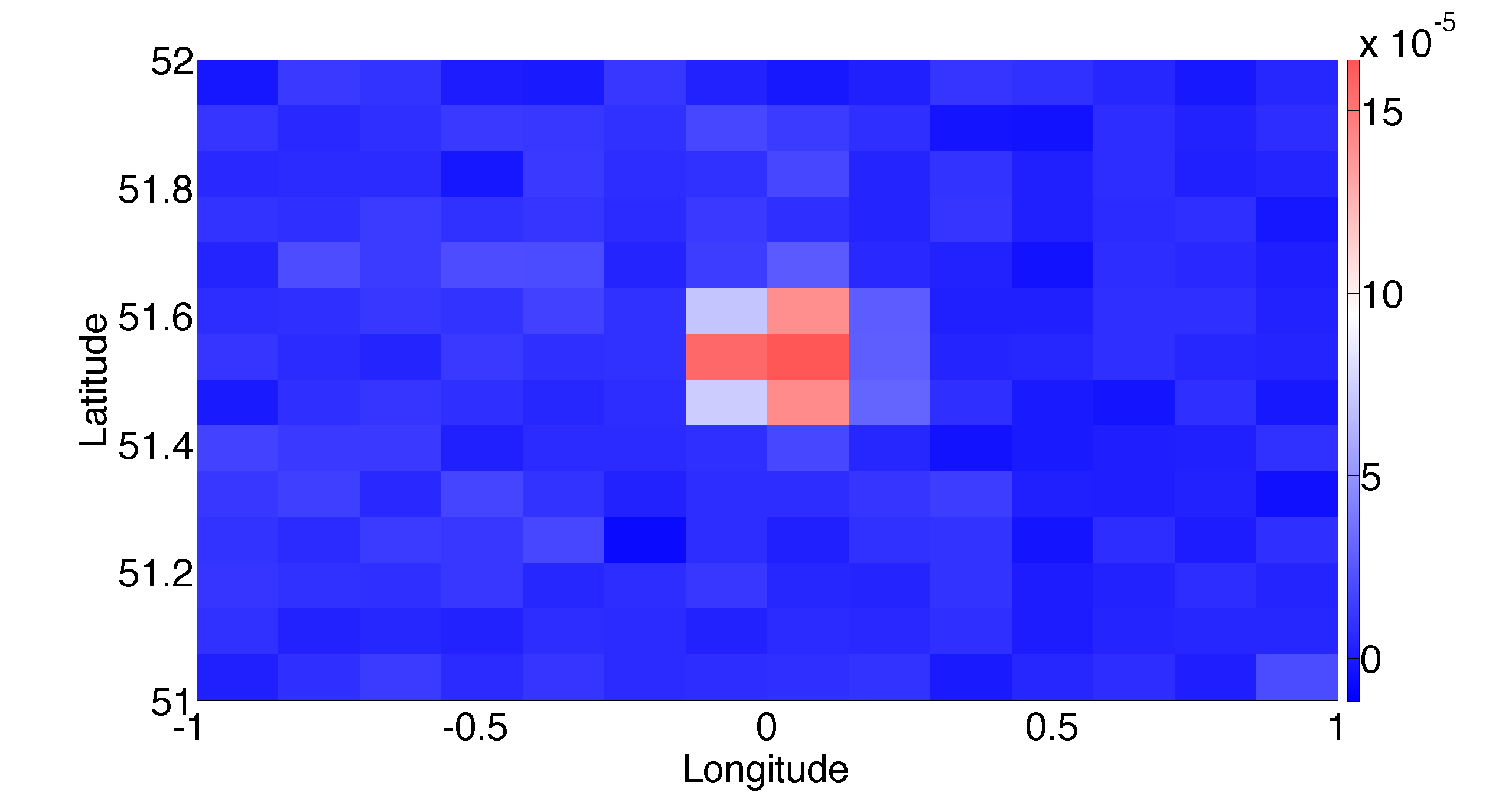}} \hspace{-0.3cm} \subfigure[\textbf{West} of grid center]{
\includegraphics[width=0.34
\textwidth,height=1.4in]{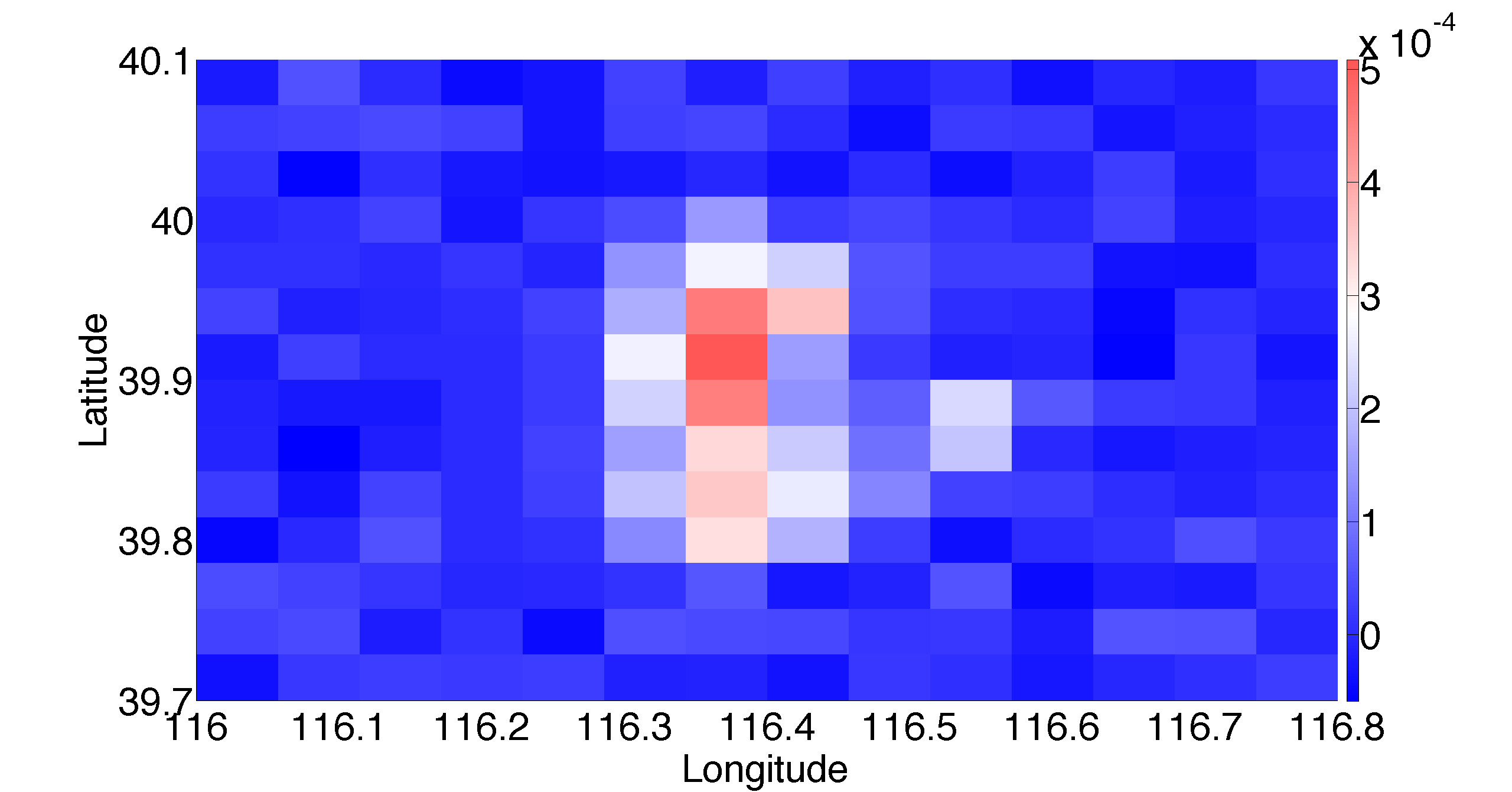}} \caption{Spatial extension of spatial relationships - probabilities of specific spatial relationships (Near, At, West) relating vertices to the center grid cell.	
\vspace{5pt}} \label{fig:qualgridcent} 
\end{figure*}

To visualize the actual models and the respective probabilities they assign to partitioned space, Figure~\ref{fig:qualgridcent} 
depicts three instances of spatial relations, with the center of the grid denoting a reference (landmark) point. Figure~\ref{fig:qualgridcent}
shows the spatial extend of relations (a) \emph{``Near''}, (b) \emph{``At''} and (c) \emph{``West''} when searching for an unknown point that is 
spatially related to the center of the grid. The examples have been derived from the New York, London, and Beijing datasets. 
The concentration of measures around qualitatively correct regions is a further indication for the correctness of our models.

\begin{figure*}[!th]
\subfigure[]{
\includegraphics[width=0.335
\textwidth,height=1.4in]{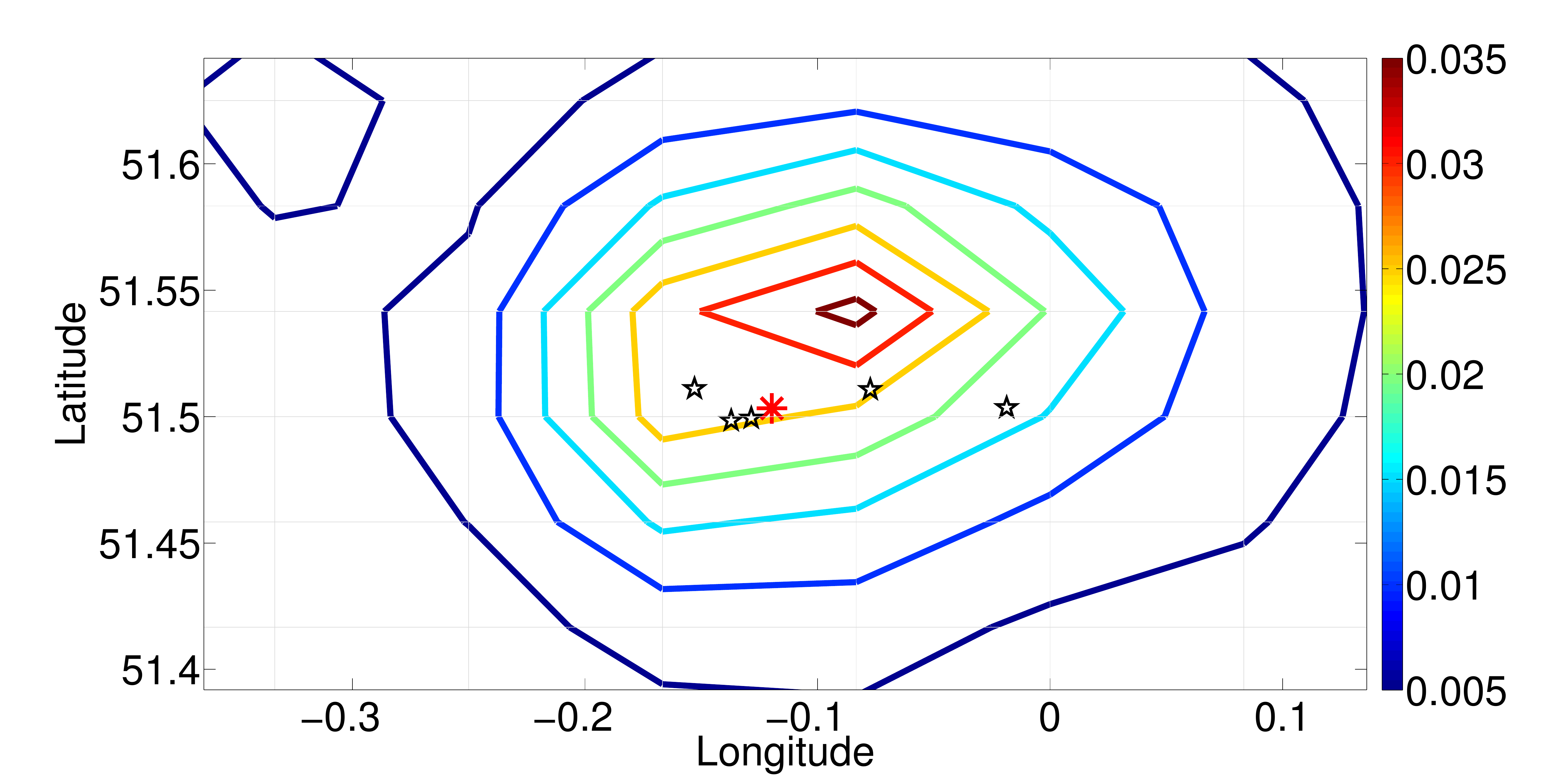}\label{subfig:londpre1}} \subfigure[]{
\includegraphics[width=0.335
\textwidth,height=1.4in]{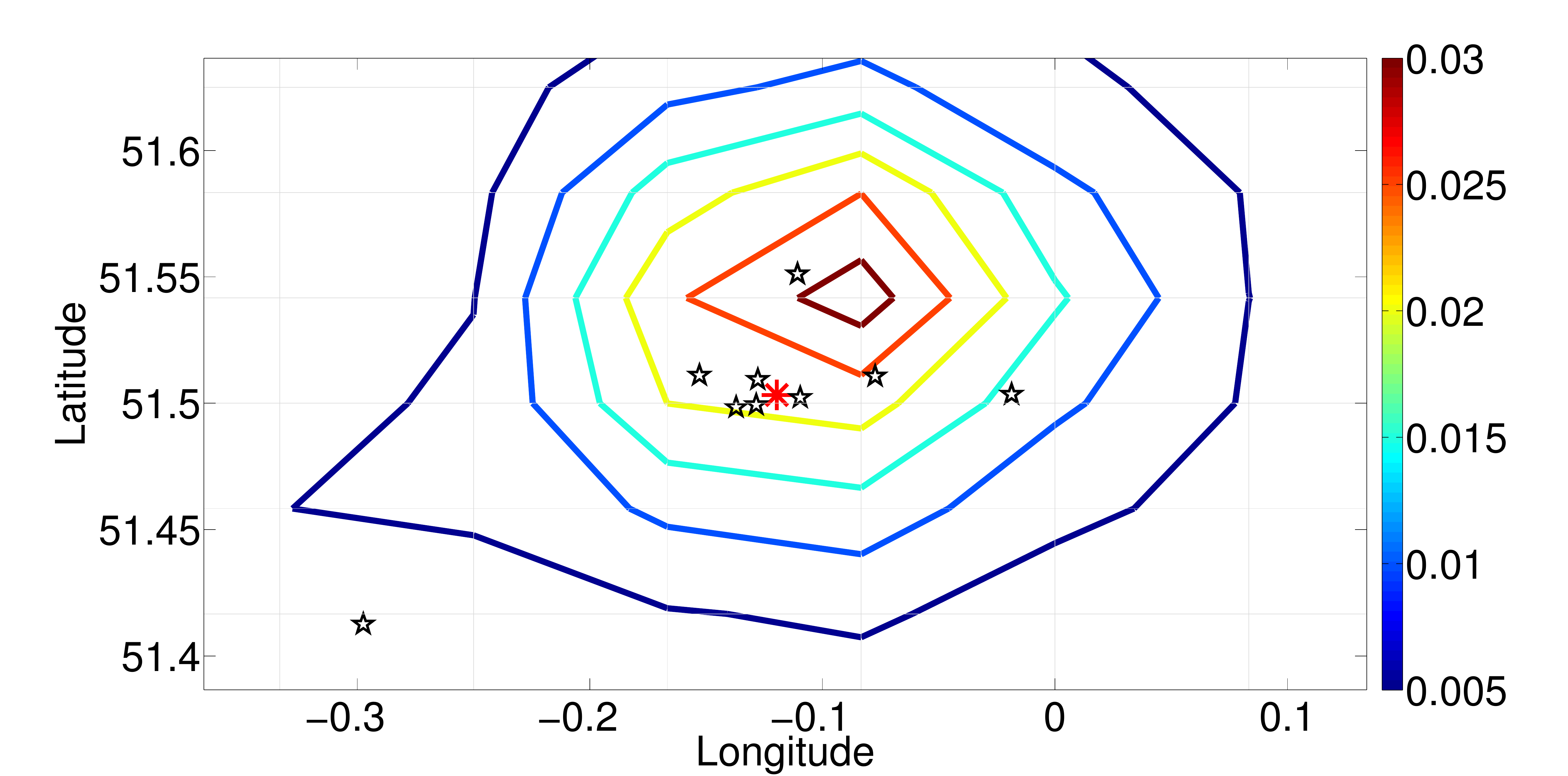}\label{subfig:londpre2}} \subfigure[]{
\includegraphics[width=0.335
\textwidth,height=1.4in]{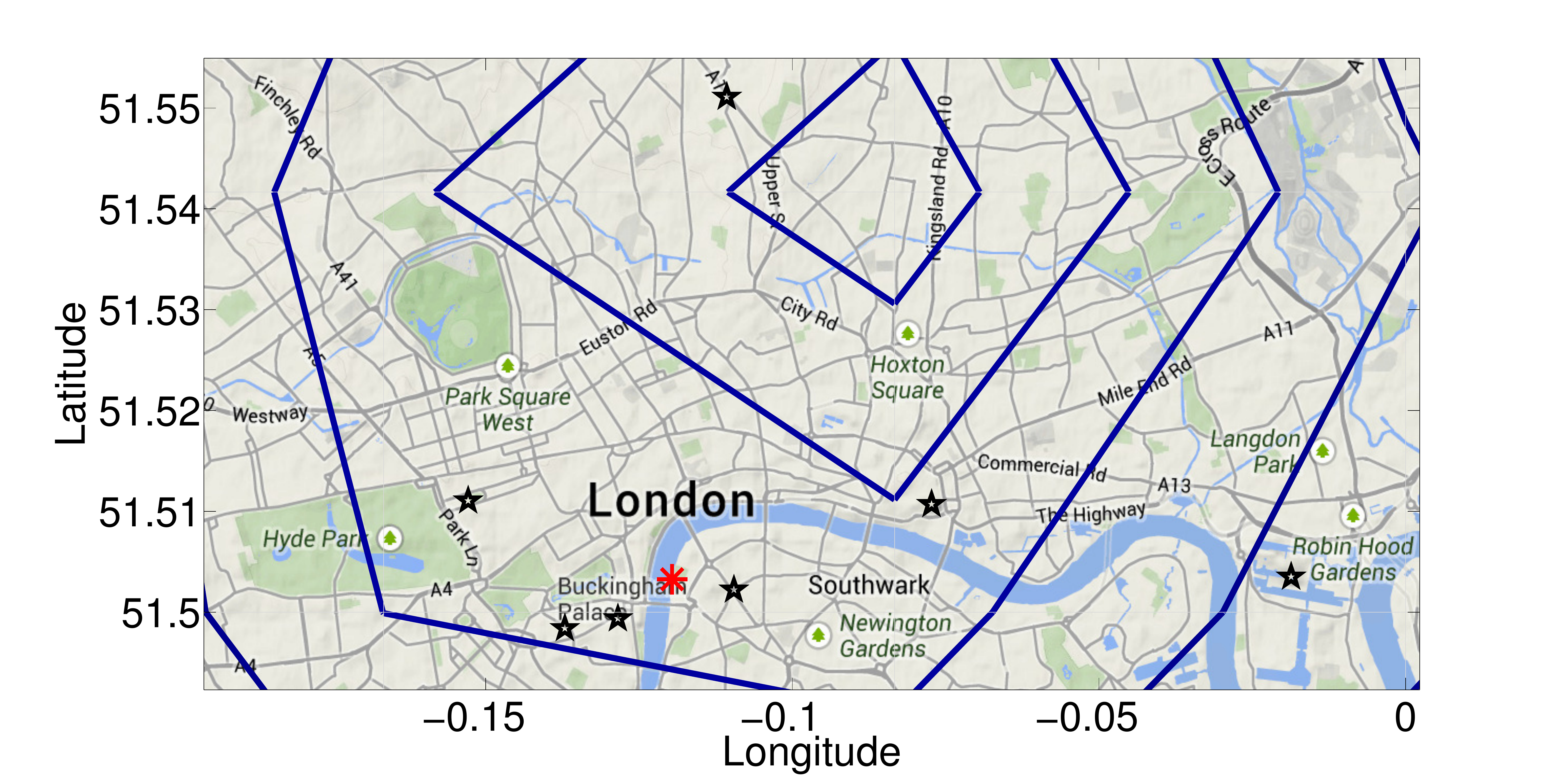}\label{subfig:londpregoog}} \\

\subfigure[]{
\includegraphics[width=0.335
\textwidth,height=1.4in]{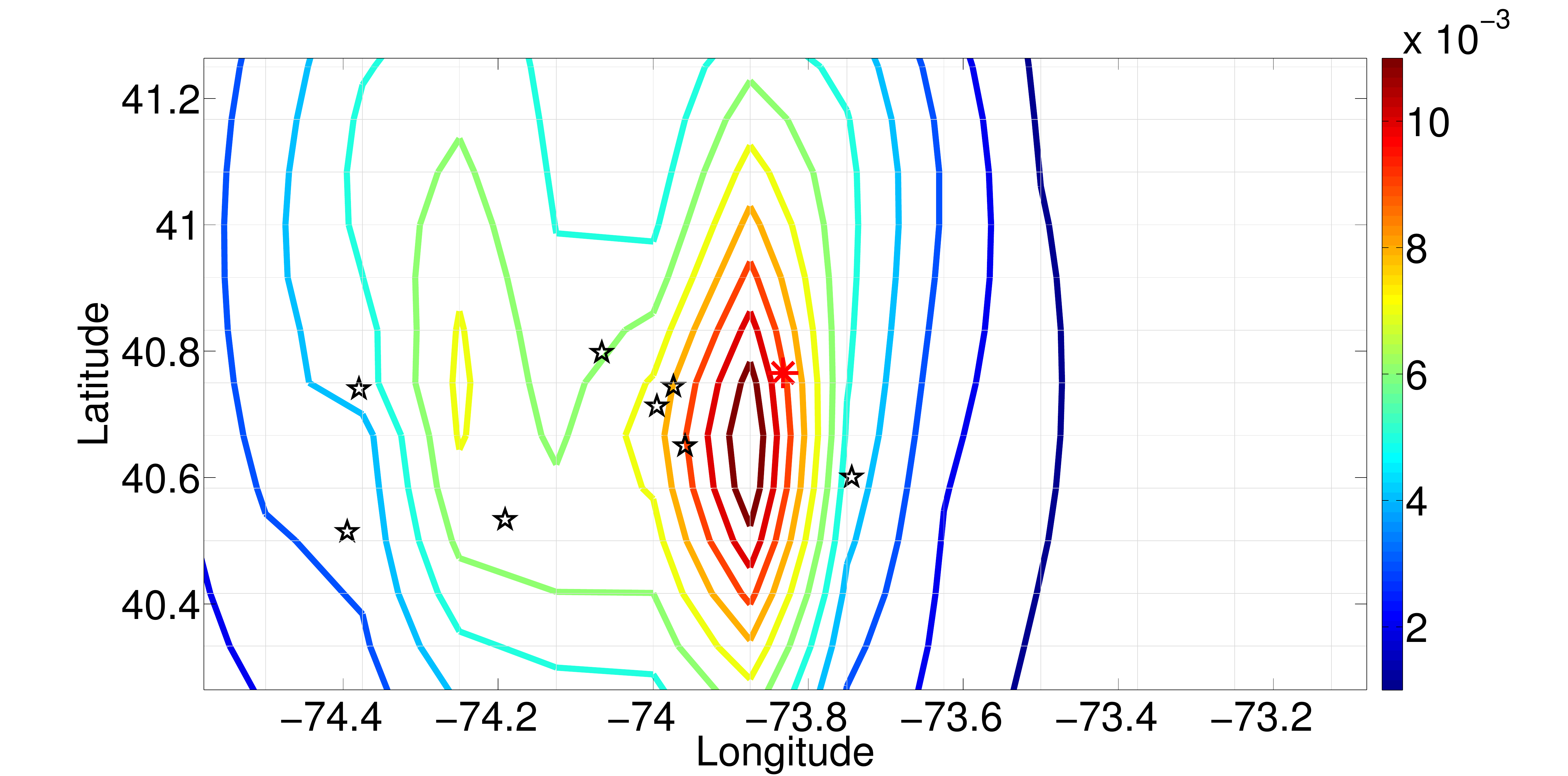}\label{subfig:nypre1}} \subfigure[]{
\includegraphics[width=0.335
\textwidth,height=1.4in]{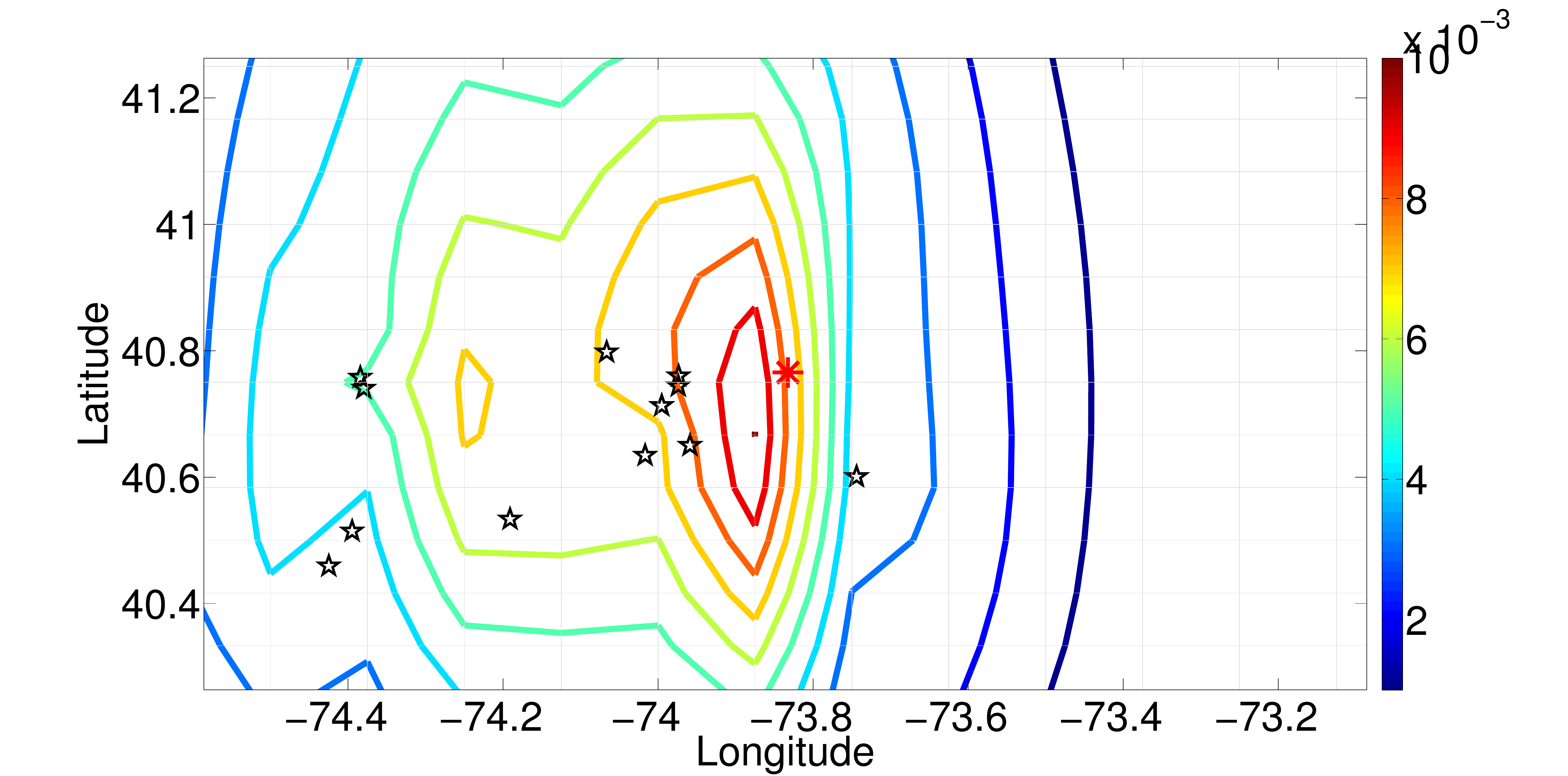}\label{subfig:nypre2}} \subfigure[]{
\includegraphics[width=0.335
\textwidth,height=1.4in]{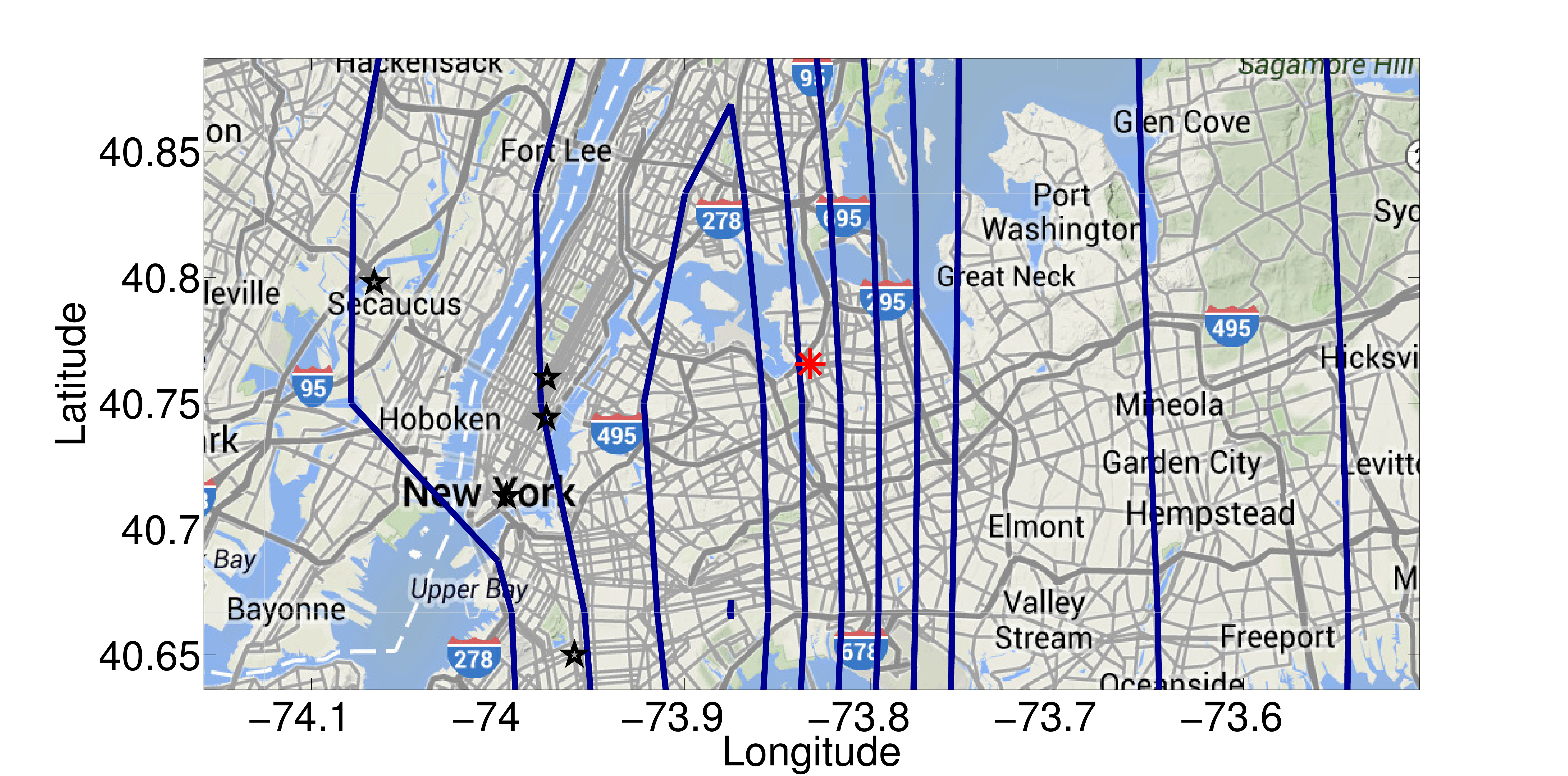}\label{subfig:nypregoog}} \\

\subfigure[]{
\includegraphics[width=0.335
\textwidth,height=1.4in]{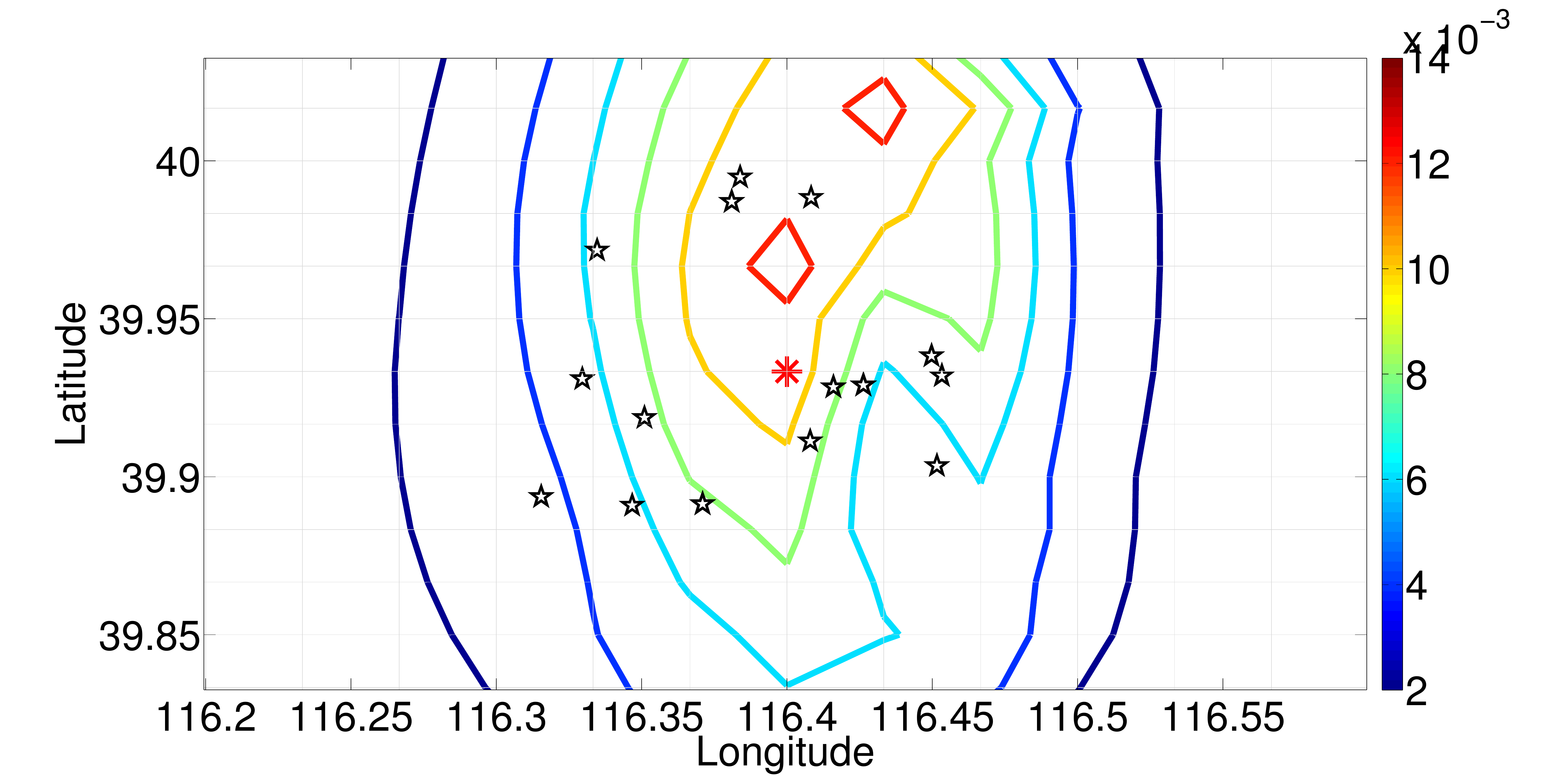}\label{subfig:beipre1}} \subfigure[]{
\includegraphics[width=0.335
\textwidth,height=1.4in]{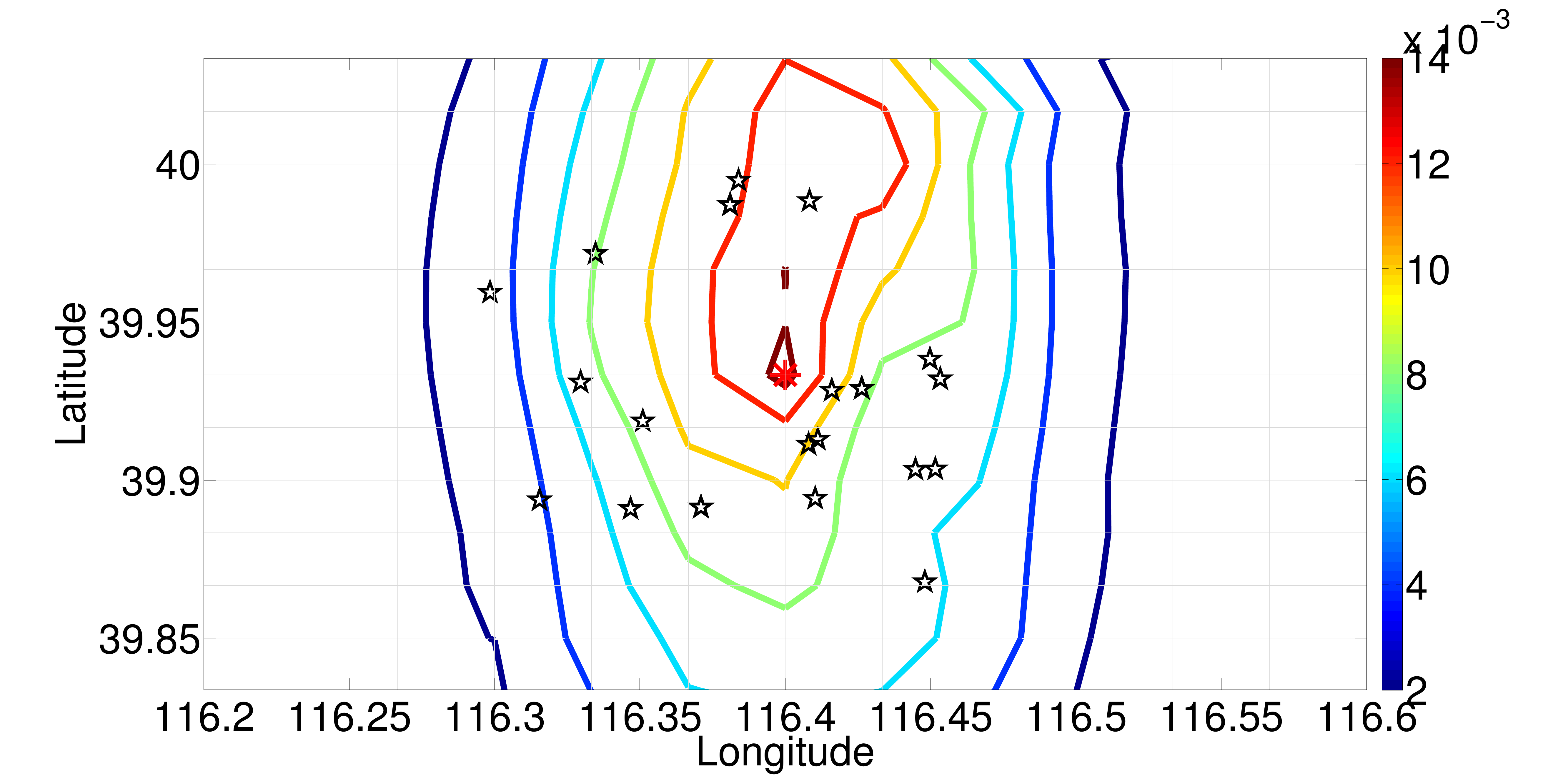}\label{subfig:beipre2}} \subfigure[]{
\includegraphics[width=0.335
\textwidth,height=1.4in]{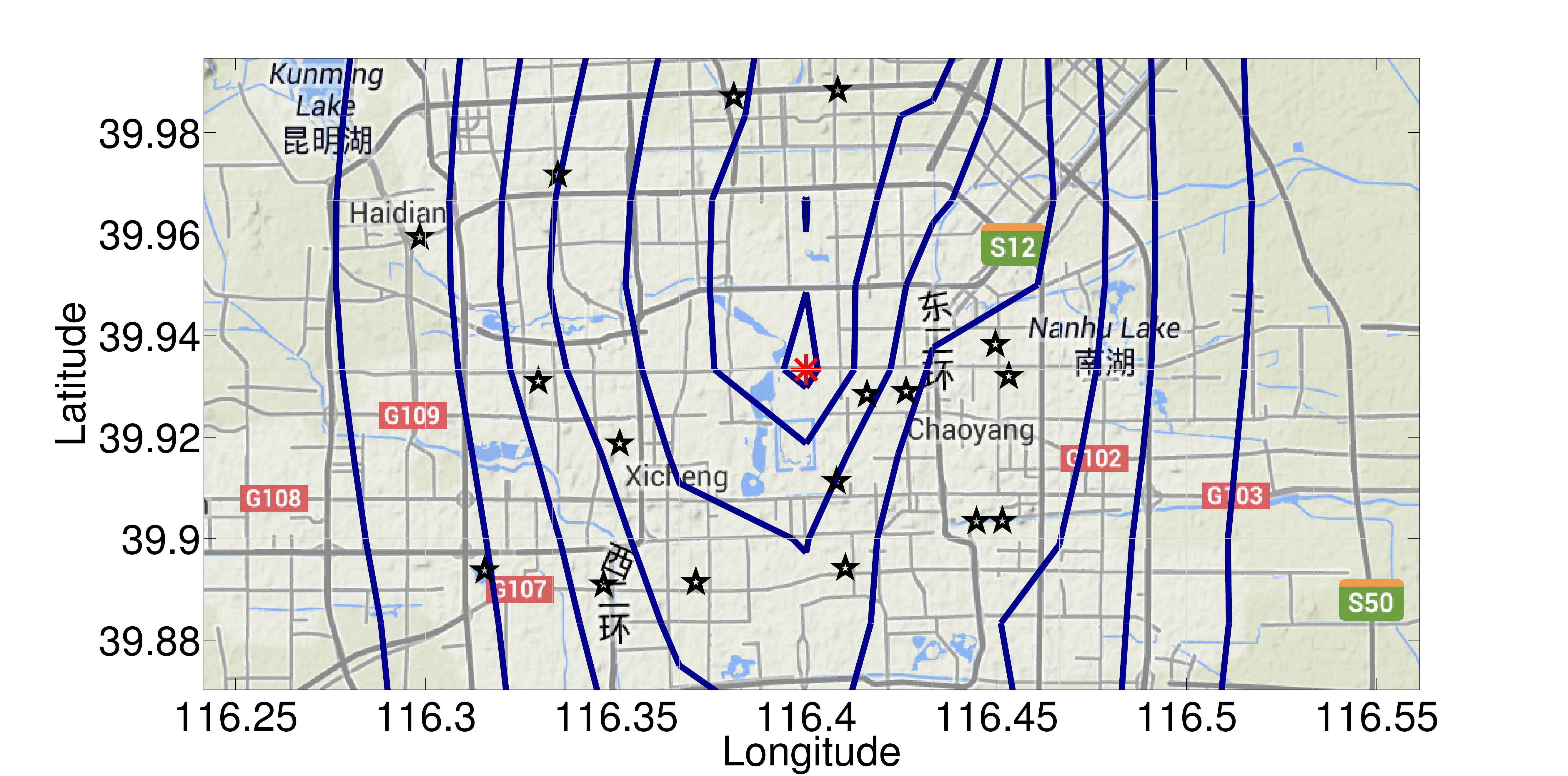}\label{subfig:beipregoog}} \\

\subfigure[]{
\includegraphics[width=0.335
\textwidth,height=1.4in]{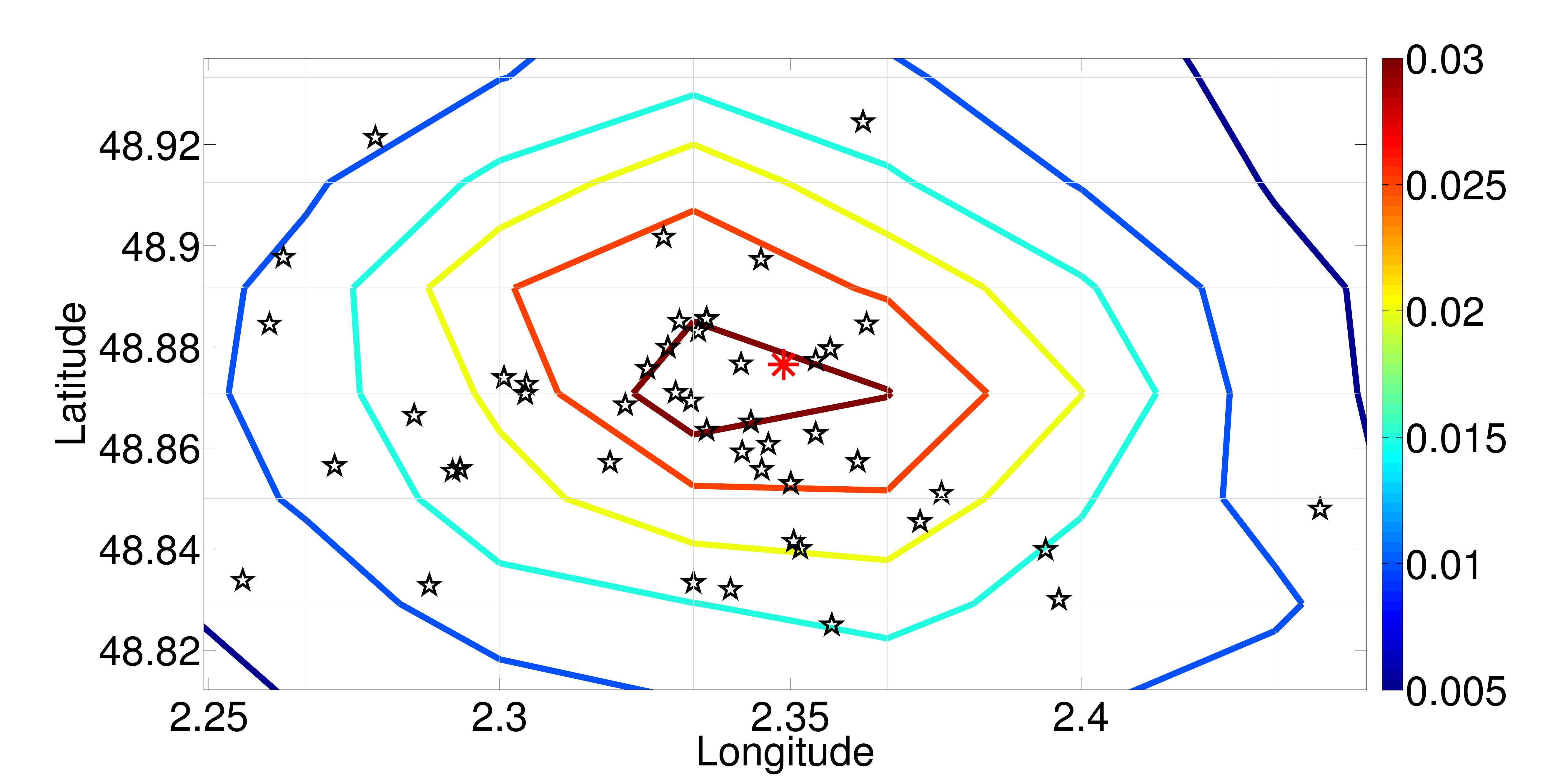}\label{subfig:papre1}} \subfigure[]{
\includegraphics[width=0.335
\textwidth,height=1.4in]{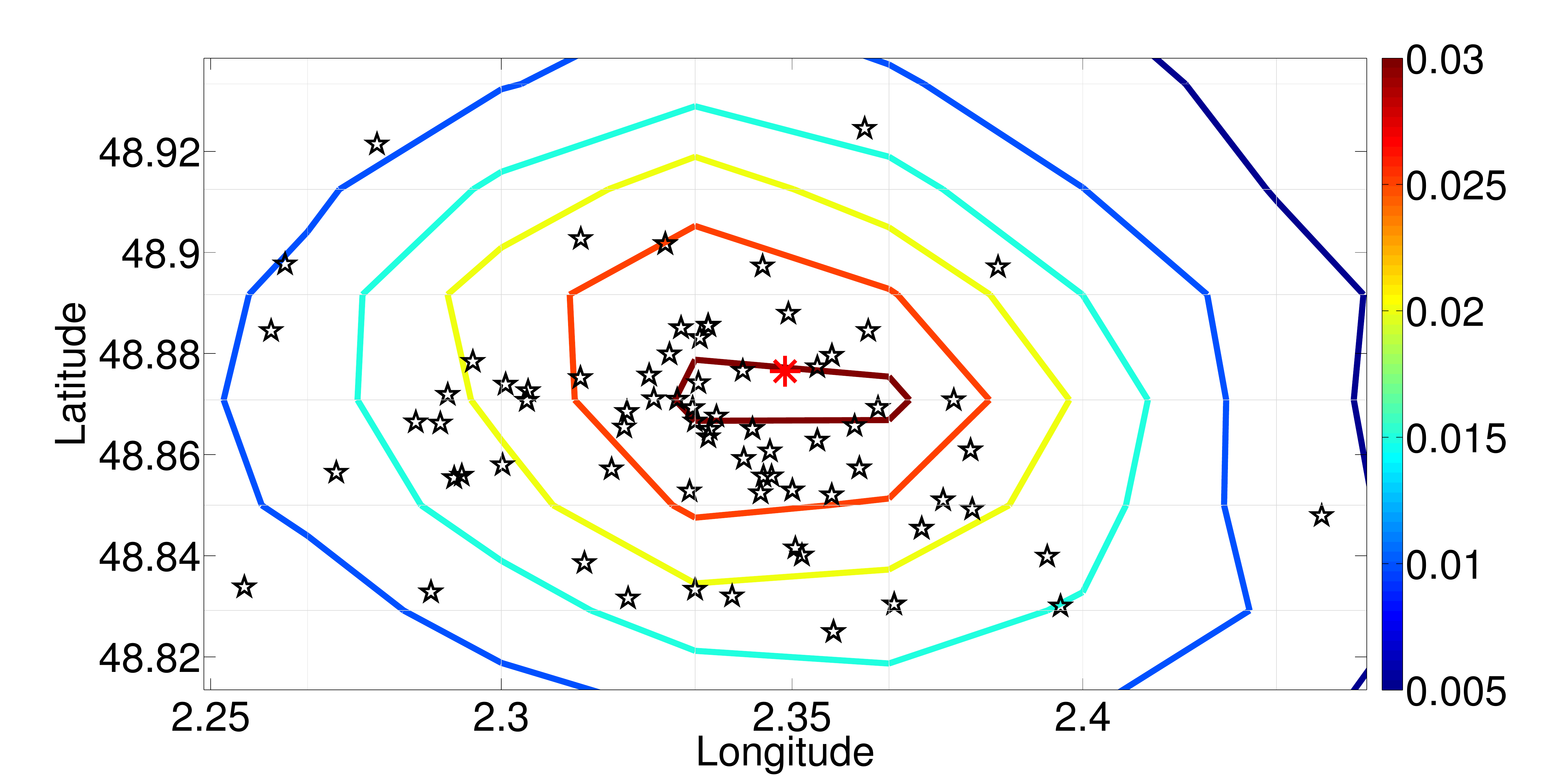}\label{subfig:papre2}} \subfigure[]{
\includegraphics[width=0.335
\textwidth,height=1.4in]{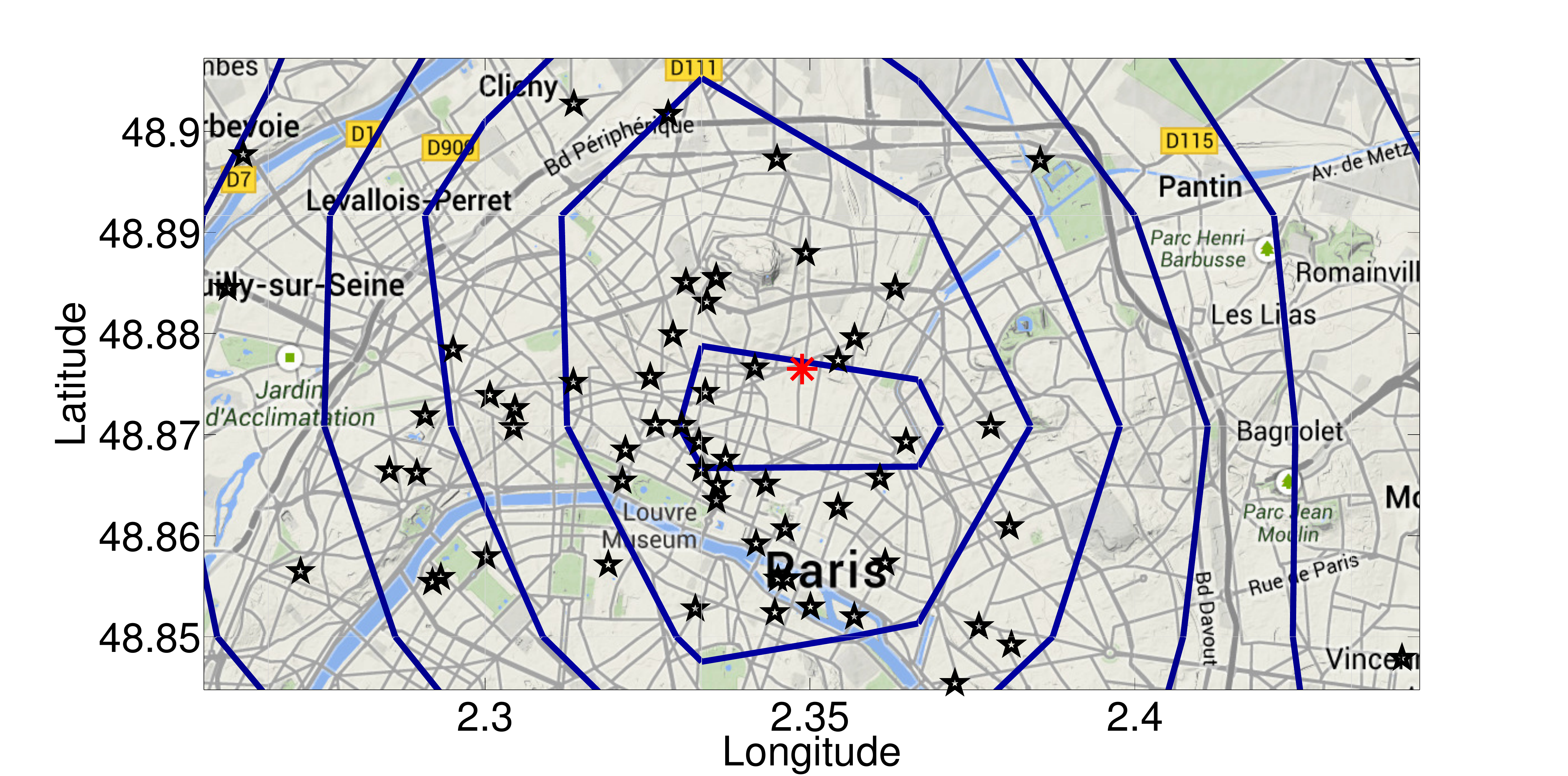}\label{subfig:papregoog}} 
 
\caption{Real world location prediction scenarios - Rows 1 to 4 are scenarios for London, New York, Beijing, 
and Paris, respectively - Columns 1-3 are a randomly selected 50\% , 100\% and 100\% (on GoogleMaps) of the discovered relations considered for prediction, respectively.\vspace{5pt}} 
\label{fig:realpred} 
\end{figure*}

\subsection{Real-world Location Estimation} \label{subsec:reallifepred}

The ultimate goal of this work is to train probabilistic models for spatial relationships so as to provide location estimation, e.g., finding the ``best pita place in Greece'', 
based on hints in the form of qualitative spatial relationships discovered in textual narratives. This is a very important application as it provides
a solution to the geocoding problem that exists on the $www$, i.e., there are millions of user referenced POIs whose coordinates do not exist in any coordinate database.

In addressing this challenge, we present four concrete location estimation scenarios. From our travel blog dataset, we extract four POIs (considered as unknown) 
whose locations are described in relation to other known POIs. Note that these cases have \textbf{not} been used in the training phase. When estimating the location of these POIs, 
we will also show the impact of an increased number of models on the quality of the location outcome. We start from a scenario where a POI is described by means of a 
few observations and subsequently increase this number. 

Figure~\ref{fig:realpred} illustrates the aforementioned scenarios. Figures~\ref{fig:realpred}(a), (b) and (c) illustrate an unknown POI (red star) in the greater area of London, 
whose position is described in relation to known POIs (black stars) using a total number of 15 spatial relations. Figure~\ref{subfig:londpre1} shows the contours of the spatial 
probability distribution when only a randomly selected $50\%$ of the spatial relations are taken into account, while Figure~\ref{subfig:londpre2} shows the final distribution considering all spatial relations. 
Finally, Figure~\ref{subfig:londpregoog} is a closeup of Figure~\ref{subfig:londpre2} with a GoogleMaps basemap overlay. 

The same approach for assessing the impact of an increased number of models is used for the cases of New York, Beijing and Paris, with a total number of 20, 70 and 200 spatial relations 
being used in each case, respectively. Again, the unknown and known POIs are marked by a red and a black star, respectively.

What should be evident from these results is the considerable prediction accuracy. 
Especially the cases of Beijing (see Figures~\ref{subfig:beipre1}~\ref{subfig:beipre2}~\ref{subfig:beipregoog}) and 
Paris (see Figures~\ref{subfig:papre1}~\ref{subfig:papre2}~\ref{subfig:papregoog}) \emph{clearly pinpoint the unknown POI location}. 
What is further encouraging is that even for the cases of London (see Figures~\ref{subfig:londpre1}~\ref{subfig:londpre2}~\ref{subfig:londpregoog}) 
and New York (see Figures~\ref{subfig:nypre1}~\ref{subfig:nypre2}~\ref{subfig:nypregoog}) where the number of relations is small, the proposed approach works reasonably well. 

As expected, the prediction accuracy increases with the number of observations (models) considered. This is confirmed by the mass of the probability moving closer 
to the unknown POI location when increasing the number of observations from a randomly selected $50\%$ (Figure~\ref{fig:realpred} $1^{st}$ column) to $100\%$ (Figure~\ref{fig:realpred} $2^{nd}$ column). 
This effect is observed for all four cases. Additionally, Table~\ref{table:distfromukn} shows the distances between the centers of the spatial probability distributions and the unknown POI locations as we increase 
the percentage of spatial relations considered for the prediction procedure. 
Here, we investigate the cases of $10\%$, $50\%$ and $100\%$ randomly selected relations. 

\begin{table}[!th]
\vspace{-8pt}
\centering
\caption{Distance between the center of the spatial probability distribution and the unknown POI.} \vspace{0.2cm}
\ra{1.1}
\begin{tabular}{l || c c c c c c c c}
\toprule
& \multicolumn{8}{c}{Percentage of relations considered} \\
\toprule
Dataset && $10\%$ &&  $50\%$ && $100\%$ && \\

\cmidrule{1-1} \cmidrule{3-3} \cmidrule{5-5} \cmidrule{7-7}
London   &&  $15.3 \text{Km}$ && $7.9\text{Km}$    && $7.7\text{Km}$   \\
New York &&  $16.2 \text{Km}$ && $11.9\text{Km}$   && $11.1\text{Km}$  \\
Beijing  &&  $14.4 \text{Km}$ && $8.6\text{Km}$    && $\textcolor[rgb]{0.4,0.1,0}{\mathbf{1.2}}\text{Km}$   \\ 
Paris    &&  $8.7 \text{Km}$  && $\textcolor[rgb]{0.4,0.1,0}{\mathbf{1.6}}\text{Km}$    && $\textcolor[rgb]{0.4,0.1,0}{\mathbf{0.8}}\text{Km}$ \\ 
\bottomrule
\end{tabular}\label{table:distfromukn}
\end{table}

The results show that as we increase the number of relations considered, we achieve more accurate estimates of the unknown POI, i.e., smaller distances between the estimated and the ground truth location. 
The improvement is considerable for all cases, with Beijing and Paris benefitting most and achieving distances of less than $2\text{Km}$ (indicated in bold in Table~\ref{table:distfromukn}). Moreover, although the estimation 
quality (accuracy as well as precision) increases with the number of observations, nevertheless, even in the case of 
a small number of observations, we can rely on the crowd as a geospatial data source for location estimation.

We can conclude that the proposed modeling using GMMs optimized by the greedy EM algorithm presented in Section~\ref{subsection:modopt} can efficiently handle the uncertainty introduced by user-contributed qualitative geospatial data. 
In combination with information extraction techniques, it provides us with the non-trivial means of textual narrative-based location estimation.

\section{Conclusions} 
\label{sec:conclusions}

The increase in available user-generated data provides a unique opportunity for the generation of rich datasets in geographical information science. With textual narrative being the most 
popular form of human expression on the internet, this work provides a method that effectively translates this text into geospatial datasets. Our specific contribution is detecting spatial 
relationships in textual narratives and using them to ``triangulate'' the position of unknown objects. This is a first step for solving the emerging geocoding problem on the internet. 
We introduce specific techniques for extracting spatial relations from textual narratives and use a novel quantitative approach based on training probabilistic models for the representation
of spatial relations. Combining these models and interpreting them as observations allows us to reason about unknown object locations. 
%
The proposed approach provides an optimized spatial relation modeling technique that achieves high-quality location estimation results as evidenced by a range of real-world datasets. 
Here, our probabilistic approach is robust with respect to handling any uncertainties that characterize geospatial observations derived from crowd-sourced textual data. 

Directions for future work include the optimization of the NLP techniques used for the automatic extraction of POIs and spatial relationship information from texts. 
Furthermore we will investigate the implementation of global prediction models, which could complement geocoding methods in our increasingly non-cartesian world. 
Also, this will enable us to evaluate additional probabilistic and deterministic modeling techniques and to develop more efficient text-to-map applications.

\bibliographystyle{abbrv}
\bibliography{LocationPrediction}


\end{document}